\documentclass[fleqn,usenatbib]{mnras}

\usepackage[T1]{fontenc}

\DeclareRobustCommand{\DEN}[3]{#2}
\let\DENthebibliography\thebibliography
\def\thebibliography{\DeclareRobustCommand{\DEN}[3]{##3}\DENthebibliography}

\usepackage[draft]{graphicx}
\usepackage{amsmath}
\usepackage{amssymb}
\usepackage[separate-uncertainty=true,detect-weight]{siunitx}
\usepackage{cleveref}
\usepackage{caption}
\usepackage{subcaption}

\usepackage{microtype}
\usepackage{booktabs}
\usepackage{threeparttable}
\usepackage{enumerate}
\usepackage{multicol}
\usepackage{multirow}

\usepackage{comment}

\usepackage{longtable}
\usepackage{threeparttablex}
\usepackage{xcolor}

\newcommand{\orcid}[1]{\href{https://orcid.org/#1}{\textcolor[HTML]{A6CE39}{\includegraphics[width=8pt]{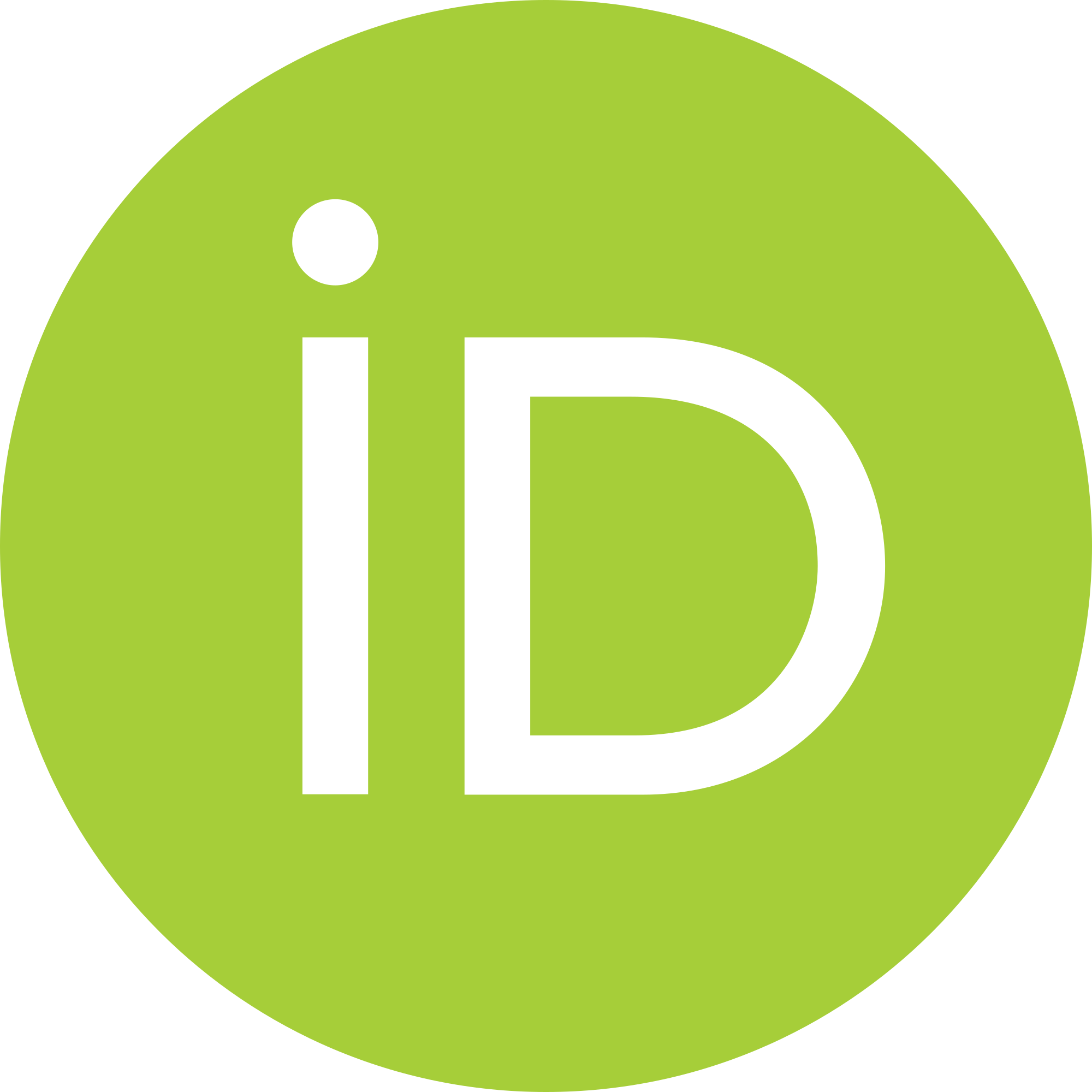}}}}

\newcommand{\hst}{\textit{HST}}
\newcommand{\hla}{\textit{HLA}}
\newcommand{\tinytim}{\textsc{TinyTim}}
\newcommand{\imfit}{\textsc{Imfit}}
\newcommand{\astrodrizzle}{\textsc{AstroDrizzle}}

\DeclareSIUnit \pc {pc}
\DeclareSIUnit \parsec {parsec}
\DeclareSIUnit \arcsec {arcsec}
\DeclareSIUnit \pixel {pixel}
\DeclareSIUnit \pixels {pixels}
\DeclareSIUnit \Msun {M_{\odot}}
\DeclareSIUnit \smass {M_{\star}}
\DeclareSIUnit \dex {dex}
\DeclareSIUnit \mag {mag}

\creflabelformat{equation}{#2#1#3}

\graphicspath{{./fig/}}

\usepackage{newtxtext,newtxmath}

\title[Photometry and Structure of New NSCs in the LV]{Photometric and Structural Parameters of Newly Discovered Nuclear Star Clusters in Local Volume Galaxies}

\author[N. Hoyer et al.]{
Nils Hoyer$^{1,2}$\thanks{Contact e-mail: \href{mailto:hoyer@mpia.de}{hoyer@mpia.de}}\orcid{0000-0001-8040-4088},
Nadine Neumayer$^{1}$\orcid{0000-0002-6922-2598},
Anil C.\ Seth$^{3}$\orcid{0000-0003-0248-5470},
Iskren Y.\ Georgiev$^{1}$\orcid{0000-0001-8471-6679}, and
Jenny E.\ Greene$^{4}$\orcid{0000-0002-5612-3427}
\medskip\\
$^{1}$Max-Planck-Institut f{\"{u}}r Astronomie, K{\"{o}}nigstuhl 17, D-69117 Heidelberg, Germany\\
$^{2}$Universit{\"{a}}t Heidelberg, Seminarstrasse 2, D-69117 Heidelberg, Germany\\
$^{3}$Department of Physics and Astronomy, University of Utah, 115 South 1400 East, Salt Lake City, UT 84112, USA\\
$^{4}$Department of Astrophysical Sciences, Princeton University, Princeton, NJ08544, USA\\
}

\date{Accepted XXX. Received YYY; in original form ZZZ}

\pubyear{2022}

\begin{document}
\label{firstpage}
\pagerange{\pageref{firstpage}--\pageref{lastpage}}
\maketitle

\begin{abstract}
  We use high-resolution \textit{Hubble Space Telescope} imaging data of dwarf galaxies in the Local Volume ($\lesssim \SI{11}{\mega\pc}$) to parameterise \num{19} newly discovered nuclear star clusters (NSCs).
  Most of the clusters have stellar masses of $M_{\star}^{\mathrm{nsc}} \lesssim \SI{e6}{\Msun}$ and compare to Galactic globular clusters in terms of ellipticity, effective radius, stellar mass, and surface density.
  The clusters are modelled with a S{\'{e}}rsic profile and their surface brightness evaluated at the effective radius reveals a tight positive correlation to the host galaxy stellar mass.
  Our data also indicate an increase in slope of the density profiles with increasing mass, perhaps indicating an increasing role for \textit{in-situ} star formation in more massive hosts.
  We evaluate the scaling relation between the clusters and their host galaxy stellar mass to find an environmental dependence:
  for NSCs in field galaxies, the slope of the relation is $\alpha = 0.82^{+0.08}_{-0.08}$ whereas $\alpha = 0.55^{+0.06}_{-0.05}$ for dwarfs in the core of the Virgo cluster.
  Restricting the fit for the cluster to $M_{\star}^{\mathrm{gal}} \geq \SI{e6.5}{\Msun}$ yields $\alpha = 0.70^{+0.08}_{-0.07}$, in agreement with the field environment within the $1 \sigma$ interval.
  The environmental dependence is due to the lowest-mass nucleated galaxies and we speculate that this is either due to an increased number of progenitor globular clusters merging to become an NSC, or due to the formation of more massive globular clusters in dense environments, depending on the initial globular cluster mass function.
  Our results clearly corroborate recent results in that there exists a tight connection between NSCs and globular clusters in dwarf galaxies.
\end{abstract}




\section{Introduction}
\label{sec:introduction}

The central regions of galaxies are interesting because of the extreme objects they host.
Besides supermassive black holes (SMBHs), which are believed to be common in high-mass galaxies \citep{kormendy2013a}, nuclear star clusters (NSCs) often occupy the centers of low- to intermediate-mass galaxies\footnote{A few galaxies are known to host both objects; see Table {3} in \citet{neumayer2020a} for a recent compilation.}.
Their size of typically a few parsecs \citep[e.g.][]{georgiev2014a,carson2015a,pechetti2020a} and high stellar mass ($M_{\star}^{\mathrm{nsc}} \sim \SI{e7}{\Msun}$, e.g.\ \citealp{georgiev2016a}) make NSCs the densest stellar systems known (see \citealp{neumayer2020a} for a review).
Similarities between these objects and globular clusters (GCs) have led to the hypothesis that NSCs are formed by the consecutive migration of GCs \citep{tremaine1975a}.
However, not all NSC properties can be explained by this formation scenario alone [e.g.\ young stellar populations in the central regions both in the Milky Way \citep[e.g.][]{lu2009a,feldmeier-krause2015a} and other nearby galaxies \citep[e.g.][]{bender2005a,seth2006a,walcher2006a,carson2015a,nguyen2017a,kacharov2018a,nguyen2019a}].
Therefore, a second formation scenario, \textit{in-situ} star formation, was proposed \citep[e.g.][]{milosavljevic2004a,agarwal2011a}.
\citet{neumayer2020a} established the idea that the relative importance of the two scenarios changes as a function of galaxy mass:
in dwarf galaxies ($M_{\star}^{\mathrm{gal}} \lesssim \SI{e9}{\Msun}$) GC migration is the dominant formation scenario, whereas in high-mass galaxies ($M_{\star}^{\mathrm{gal}} \gtrsim \SI{e9}{\Msun}$) the majority of the NSC stellar mass is build-up \textit{in-situ}.
Most recently, this transition was observed in dwarf early-type galaxies \citep{fahrion2020a,fahrion2021a}.
In addition, using the theoretical framework of \citet{leaman2022a} of the build-up of NSCs through GC migration, \citet{fahrion2022a} quantified the \textit{in-situ} fraction of NSCs which appears to decline towards low NSC masses.

NSC occurrence is not uniform and varies with host galaxy stellar mass, morphological type, and environment.
It is now well established that NSCs are most common in galaxies with stellar masses of $M_{\star}^{\mathrm{gal}} \sim \SI{e9.5}{\Msun}$ \citep{sanchez-janssen2019a,neumayer2020a,hoyer2021a} and that their rate of occurrence declines towards lower and higher stellar mass.
It is speculated that the rivalry between SMBHs and NSCs at the high-mass end can lead to the evaporation of the cluster due to tidal heating \citep[e.g.][]{cote2006a} and binary black hole mergers \citep[e.g.][]{antonini2015b}.
At the low-mass end, it seems that NSCs and GCs are closely linked \citep[e.g.][]{sanchez-janssen2019a,carlsten2022a} and that the lack of GCs in lower mass galaxies drives the declining NSC frequency.

Numerous new detections were made over the last few years with ground-based surveys, increasing the total number of NSCs beyond \num{1000} \citep{munoz2015a,venhola2018a,sanchez-janssen2019a,habas2020a,carlsten2020a,su2021a,poulain2021a}.
While ground-based surveys have the clear advantage of rapidly increasing number statistics, with the exception of the closest systems, their data cannot be used to determine structural parameters, and very few structural parameter estimates are available for NSCs in low-mass galaxies.
To investigate this parameter space, high-resolution imaging data are required, as provided by the \textit{Hubble Space Telescope} ({\hst}).
In the past, numerous studies have used {\hst} data to analyse NSCs \citep[e.g.][]{carollo1998b,boeker1999b,boeker2002a,walcher2005a,cote2006a,seth2006a,baldassare2014a,georgiev2014a,pechetti2020a}, even in galaxies in the \SI{\sim 100}{\mega\pc} distant Coma galaxy cluster \citep{denbrok2014a,zanatta2021a}.

Recently, we analysed {\hst} data for more than \num{600} galaxies to constrain the frequency of NSCs in the Local Volume \citep{hoyer2021a}.
During this analysis, we discovered \num{21} new NSCs that had not been previously catalogued.
In this paper we present structural parameter measurements of these \num{21} newly discovered NSCs.
We investigate possible relations of the NSCs' parameters and their connection to the underlying host galaxy.

\Cref{sec:identification_of_nuclear_star_clusters} briefly introduces the data and describes the method of identifying nucleated galaxies.
Details regarding image processing, PSF generation, and the fitting procedure are presented in \Cref{sec:analysis}.
\Cref{sec:results_and_discussion} discusses our results on NSC parameters, their wavelength dependence, and scaling relations.
We conclude in \Cref{sec:conclusions}.
Additional remarks regarding uncertainties are given in \Cref{sec:assessing_uncertainties}.
All data tables are presented in \Cref{sec:data_tables} and are also available online in a machine-readable format.

\section{Identification of Nuclear Star Clusters}
\label{sec:identification_of_nuclear_star_clusters}

In \citet{hoyer2021a} we determined if galaxies have NSCs through a multi-step process using {\hst} ACS, WFPC2, and WFC3 data.
In a first step, we visually inspected all available imaging data.
During this step, we removed galaxies with obscured centres or if their centres were not visible on the data.
Furthermore, we identified bright central and compact objects as potential NSCs.

\begin{figure*}
  \centering
  \includegraphics[width=0.33\textwidth]{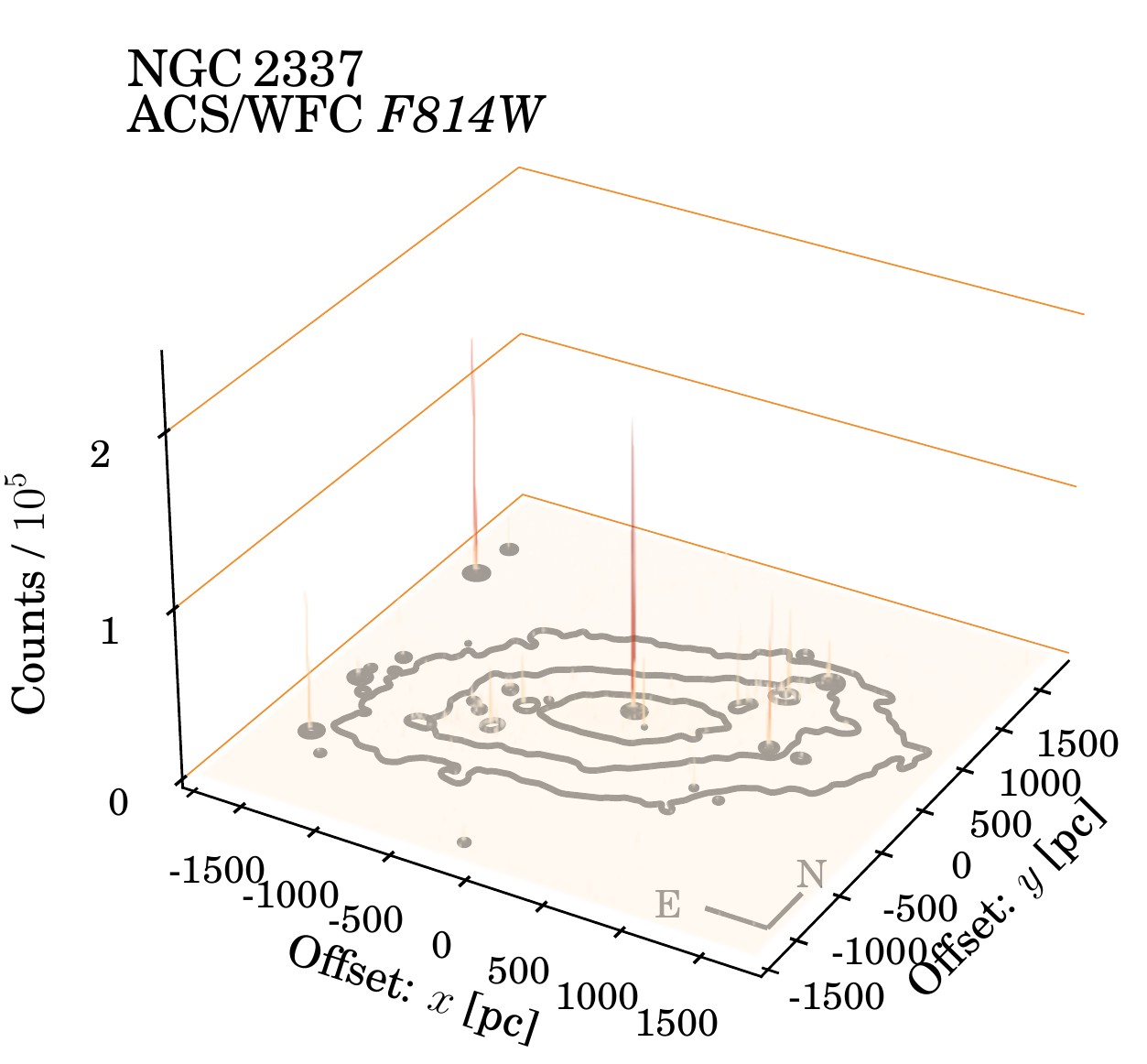} \hfill
  \includegraphics[width=0.33\textwidth]{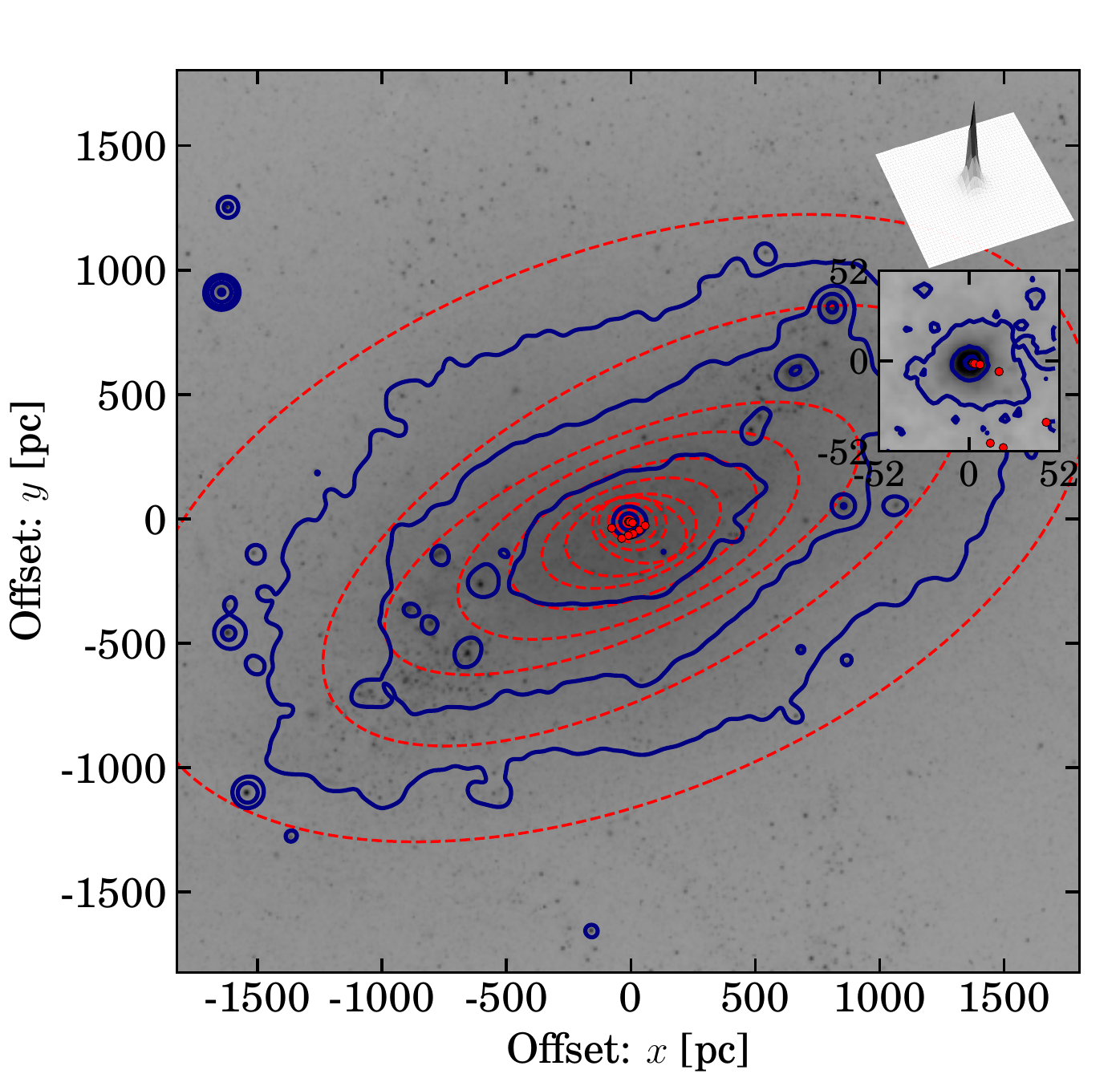} \hfill
  \includegraphics[width=0.33\textwidth]{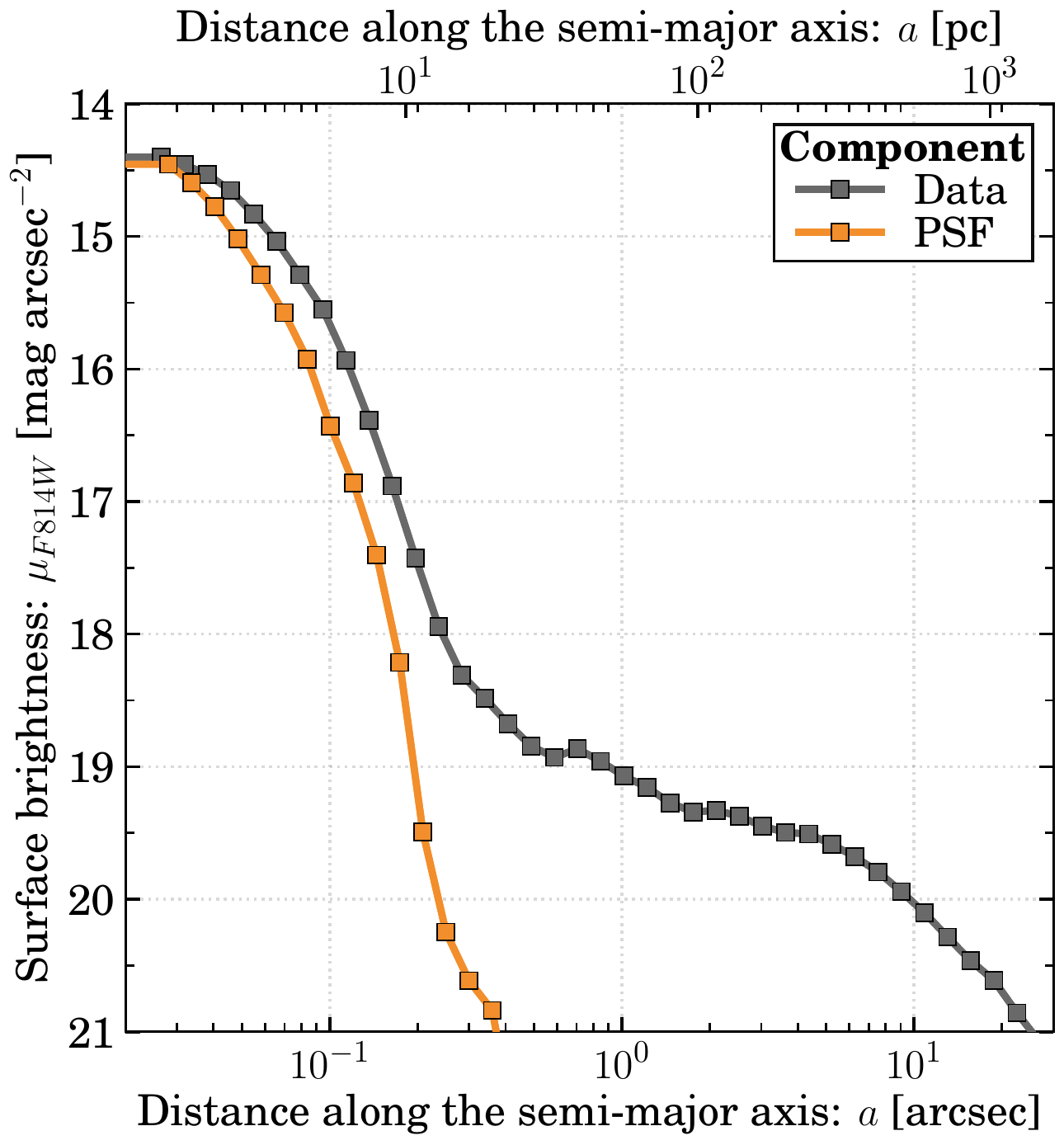}
  \caption{%
    Nuclear star cluster (NSC) identification procedure using NGC{\,}2337 as an example, based on the {\hst} ACS/WFC $F814W$ data product.
    \textit{Left panel}:
    Three-dimensional contour plot of NGC{\,}2337 centered on the NSC.
    North is up and East is left.
    Contour lines represent the profile of a smoothed version of the data.
    For smoothing, we use a Gaussian kernel with standard deviation of five pixels.
    The NSC is the brightest source connected to the galaxy.
    \textit{Middle panel}:
    Two-dimensional contour plot of the same data as in the left panel keeping the same orientation and blue contour lines.
    Red dashed lines give elliptical isophotes which were fit to the smoothed data; red points mark their centres.
    The two top right panels highlight the direct vicinity of the NSC.
    \textit{Right panel}:
    One-dimensional surface brightness ($\mu_{F814W}$) plot of the elliptical isophotes (red colour, middle panel) as a function of semi-major axis ($a$) in both arcsecs and parsecs.
    Gray squares indicate the data.
    Orange squares show the extent of a {\tinytim}-generated PSF for the position of the NSC on the chip.
  }
  \label{fig:imaging_example}
\end{figure*}

Next, we created multiple three- and two-dimensional figures, as well as a one-dimensional surface brightness plot.
The aim of these plots is to (1) indicate the intensity of the compact source compared to its host galaxy, (2) check the position of the compact source within the galactic body, and (3) visually inspect the extent of the compact source and its host galaxy.
As an example, \Cref{fig:imaging_example} shows these plots for the ACS/WFC $F814W$ data of NGC{\,}2337, the most massive galaxy in our sample of newly discovered NSCs.
Given that NSCs are dense stellar systems close to the photometric and kinematic centres of their hosts \citep{neumayer2011a,poulain2021a}, we expect them to (1) have the highest intensity within the galactic body and (2) lie `close' to the centres of elliptical isophotes which were fit to the galactic body, as visualised in the middle panel of \Cref{fig:imaging_example}.
In this step, we removed potential NSC candidates if they lay in the outskirts of their host galaxy (with typical distances of \SI{\geq 1}{\kilo\pc} to the galactic centre) or if several other compact sources had similar intensities, indicating that the compact source is either a faint foreground star or one of many GCs.

\begin{figure*}
  \centering
  \includegraphics[width=0.85\textwidth]{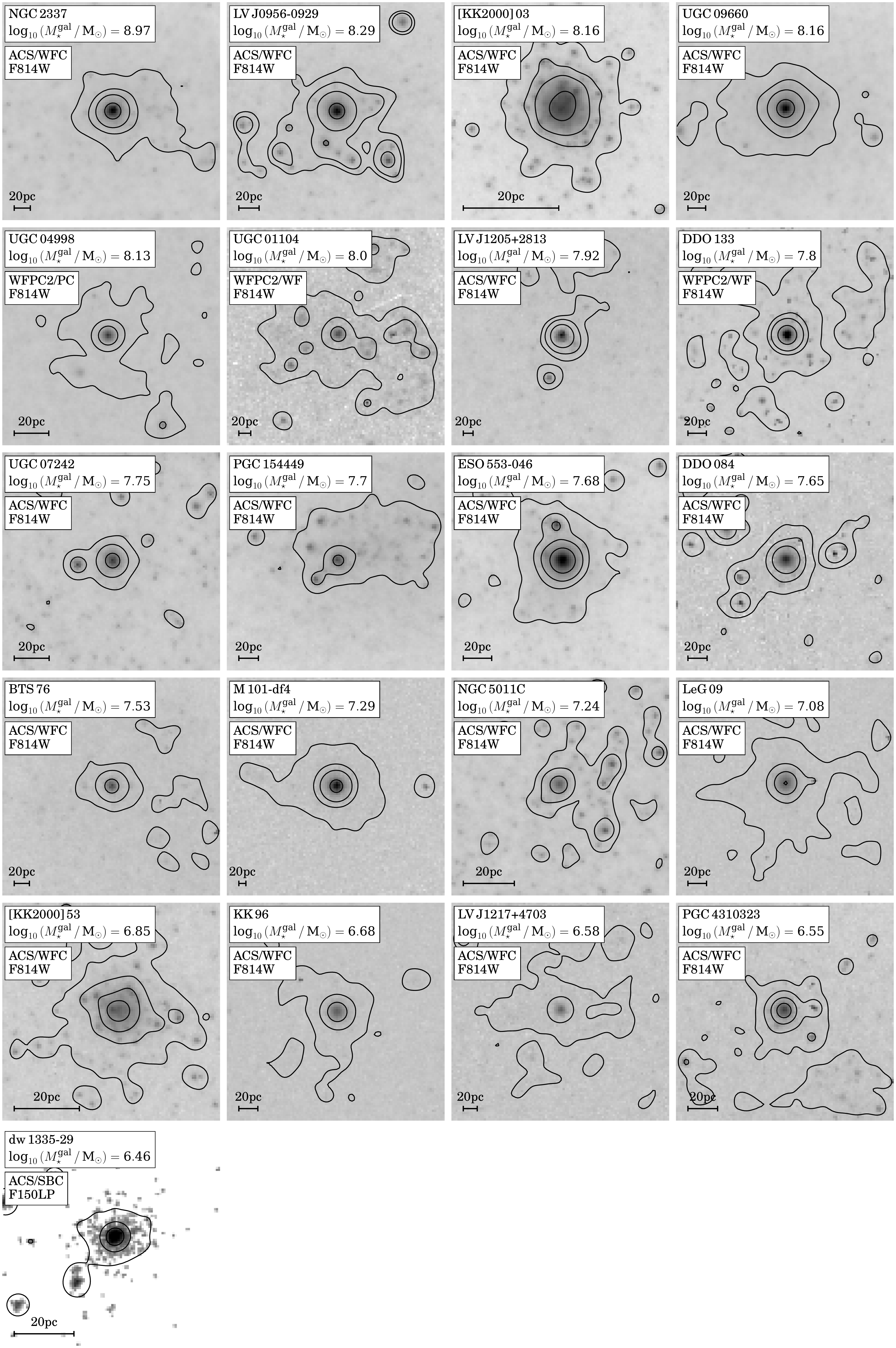}
  \caption{%
    Collage of the \num{21} newly discovered nuclear star clusters, sorted by host galaxy stellar mass from top left to bottom right.
    Each image shows a square box of side length \SI{100}{\pixels} centered on the nucleus; \SI{20}{\parsec} at the distance to the galaxies are indicated in each panel.
    North is up and East is left.
    The contour lines were derived from a smoothed version of the data using a Gaussian kernel with standard deviation of three pixels.
  }
  \label{fig:nsc_snippets}
\end{figure*}

In a third step we performed a two-dimensional fit to the data to extract the magnitude and extent of the compact source.
A point spread function (PSF) was generated at the location of the compact object of the chip using {\tinytim} \citep{krist1993a,krist1995a} and the fit was performed with {\imfit} \citep{erwin2015a}.
The PSF was then convolved with a S{\'{e}}rsic profile \citep{sersic1968a} of the form
\begin{equation}
  \label{equ:sersic_profile}
  I(r) = I_{\mathrm{eff}} \exp \bigg \{ -b_{n} \bigg [ \bigg ( \frac{r}{r_{\mathrm{eff}}} \bigg )^{1 / n} - 1 \bigg ] \bigg \}\quad,
\end{equation}
and fit to the data.
Here $r_{\mathrm{eff}}$ is the effective radius, $I_{\mathrm{eff}}$ the intensity at the effective radius, $n$ the S{\'{e}}rsic index, and $b_{n}$ solves $\Gamma(2n) = 2\gamma(2n, b_{n})$ where $\gamma(x, a)$ is the incomplete and $\Gamma(x)$ the usual Gamma function \citep[see also][]{graham2005a}.
Such S{\'{e}}rsic profiles have been widely used in fitting nearby NSCs in the recent literature \citep[e.g.][]{graham2009b,carson2015a,pechetti2020a}.
If the extent of the compact source was larger than \SI{20}{\percent} of the width of the PSF (typically \SI{\geq 1}{\pc}), we classified the compact object as an NSC and considered it for further analysis.
In total, \num{21} compact objects in the central regions of galaxies fulfilled all requirements (including NGC{\,}2337 in \Cref{fig:imaging_example}), were classified as NSCs, and are new detections.
We show images of these \num{21} objects in \Cref{fig:nsc_snippets}.

\section{Analysis}
\label{sec:analysis}

\subsection{Image processing}
\label{subsec:image_processing}

For each galaxy, we combined single exposures using flat-fielded data products obtained from the \textit{Hubble Legacy Archive}\footnote{\url{http://hla.stsci.edu/}} ({\hla}).
Instead of using the available final data products, we prefer to combine the exposures ourselves to ensure a homogeneous calibration process and to control the pixel scale of the drizzle output.

In a first step, we obtained the raw exposures from the ACS, WFPC2, and WFC3 instruments and updated the world coordinate system of each exposure using the latest reference files.
This step was required to obtain a subpixel accuracy between individual exposures and to avoid a systematic broadening of the NSC.
We fed the aligned exposures to {\astrodrizzle} \citep{gonzaga2012a} which combined them into a single science product.
No sky subtraction was performed.
The program allows the user to modify the pixel fraction and pixel scale of the final drizzled image.
The pixel fraction varies between zero and one where a value of zero corresponds to pure interlacing and a value of one to shifting and addition of pixel values from individual exposures.
The drizzle algorithm \citep{fruchter1997a} combines both techniques and enables a gain in image resolution and reduction in correlated noise.
We chose a value of \num{0.75} for the pixel fraction which is the smallest value for which no artifacts appeared in the weight map of the output image of {\astrodrizzle}.
Increasing the value towards one did not change the fit results.

In addition to the pixel fraction, we changed the pixel scale for the ACS data products.
The image resolution of the WFPC2 and WFC3 products remain unchanged.
The limiting factor in increasing the spatial resolution of our ACS data products is given by the extent of the core of the theoretical PSF.
This value is presented for the \textit{F550M} band in the ACS manual\footnote{See \citet{acs_manual} and \url{https://www.stsci.edu/hst/instrumentation/acs}.}.
We determine the width of the PSF in a different filter by constructing the ratio of the full-width-half-maximum between {\tinytim}-generated PSFs in that filter and the \textit{F550M} band.
The pixel scale of the ACS images was chosen to Nyquist sample the PSF full-width-half-maximum at each wavelength.
It ranges between \SI{0.0415}{\arcsec\per\pixel} and \SI{0.0472}{\arcsec\per\pixel}, depending on the filter.
The final resolution of each data product is indicated in \Cref{tab:available_data} in \Cref{sec:data_tables}.

Finally, to perform the actual fit, we selected a square region of side length 100 pixels around the position of the NSC.
Depending on the image resolution and the distance to the galaxy, this square region covers an area between $\SI{\sim 50}{\parsec} \times \SI{\sim 50}{\parsec}$ and $\SI{\sim 500}{\parsec} \times \SI{\sim 500}{\parsec}$.
As NSCs typically have effective radii of a few parsecs \citep[e.g.][]{neumayer2020a} the selected area ensures that the wings of the NSCs are well captured.
Nevertheless, we verified that both doubling the side length of the square and reducing it down to 60 pixels does not affect the final results.

\subsection{PSF generation}
\label{subsec:psf_generation}

Detailed knowledge of the PSF at the location of the NSC is required to reliably measure effective radii as they are generally compact and cover only a few pixels on the exposure.
The PSF can be recovered from stellar sources in the image or generated synthetically.
We decide to generate synthetic PSFs using {\tinytim} for three reasons:
\begin{itemize}
  \item[(i)] It is difficult to find non-saturated stars in the proximity of the NSC.
  Stars far away from the NSC should not be used as the {\hst} PSFs vary significantly across the chip.
  \item[(ii)] The extracted PSF from stars may be subject to variations due to the positions of the stars on the chip and their stellar type.
  \item[(iii)] Extracting a PSF from stars results in an inhomogeneous treatment of using PSFs across the whole NSC sample.
\end{itemize}
Synthetic PSFs avoid these issues and allow us to control the input parameters such as position on the chip and the assumed stellar type.

To generate a PSF, we first determined the position of the NSC on each exposure.
PSFs were generated using {\tinytim} and the location of the NSC on the chips, while assuming a \texttt{G2V} spectral type ($V - I = \SI{0.71}{\mag}$) for the artificial star.
After the PSF generation, we created a copy of the science exposures and subtract the image data from the first header file.
The PSF was then added to the flattened image data at the previous location of the NSC.
We then fed the data to {\astrodrizzle} and executed the program with the same settings as for the science data.
This step ensures that the final PSF, which was extracted from the output of the program, is processed in the same way as the NSC on the science data.

Note that the inclusion of the {\astrodrizzle} processing step is crucial as the resulting PSF will change depending on the chosen parameter settings.
In our tests the core of the resulting PSFs were slightly larger than the core of any of the {\tinytim}-generated PSFs.
Therefore, not performing this step results in systematically larger effective radii compared to their `true' values.
We discuss this effect and other potential systematic uncertainties, such as the spectral type of the artificial star or the uncertainty on positioning the PSFs on the chips in \Cref{subsubsec:systematic_uncertainties}.

\subsection{Fitting procedure}
\label{subsec:fitting_procedure}

We assume that the NSC light distribution can be accurately modelled with a single S{\'{e}}rsic profile \citep{sersic1968a}, as is common practice in the literature \citep[e.g.][]{turner2012a,baldassare2014a,carson2015a,pechetti2020a}.
For the background light, which includes the galaxy itself, we used a flat background assuming that local flatness holds in the proximity of the NSC.
The only two exceptions are UGC{\,}01104 and UGC{\,}09660 where the fit required a second S{\'{e}}rsic profile for the underlying galaxy\footnote{If only a single S{\'{e}}rsic profile is used the fit `prefers' to fit the underlying profile over the NSC.}.
Using version 1.8.0.\ of {\imfit}, the S{\'{e}}rsic profile was convolved with the PSF and fit to the data where the goodness of fit is evaluated via standard $\chi^{2}$ statistics.
The data were fit using a differential evolution solver with Latin hypercube sampling \citep{storn1997a}.
The solver is less prone to be stuck in local minima compared to other solvers available in {\imfit} and does not rely on initial parameter estimates as parameter values are randomly sampled given lower and upper boundaries.
We list the chosen boundary values in \Cref{tab:parameter_boundaries} and note that they are kept the same for all NSCs in all filters.

We additionally tested that other model functions do not significantly change the resulting parameter estimates.
For the NSC, the tests included a King profile, multiple S{\'{e}}rsic profiles, point sources, nuclear rings, and various combinations.
According to the Bayesian Information Criteria, none of these fits significantly improved over a fit with a single S{\'{e}}rsic profile.
In addition, by adding a S{\'{e}}rsic profile to the flat background component to account for the underlying galaxy, we found that the assumption of local flatness is justified.
We verified that using Cash statistics instead of the classical $\chi^{2}$ statistics does not change the results.
We defer to \Cref{subsubsec:model_functions} for a detailed discussion regarding the choice and justification of these models.

For each NSC, the fits in different filters were performed independently of each other.
However, in some cases the S{\'{e}}rsic index diverged towards the upper boundary in one filter, but not in the other.
In these cases (BTS{\,}76, DDO{\,}084, ESO{\,}553-046, [KK2000]{\,}53, KK{\,}96, LeG{\,}09, LV{\,}J1217+4703, NGC{\,}5011C), we kept all structural parameters of the fit with the diverging S{\'{e}}rsic index fixed such that only the $(x,y)$ position, the intensity at the effective radius, and the flat background component were allowed to vary.

For a number of galaxies the S{\'{e}}rsic index diverged towards high values in all available filters.
This behaviour persisted when considering a single point source or a point source in combination with a S{\'{e}}rsic profile, and also occured independently of the settings chosen for {\astrodrizzle}, {\tinytim}, and {\imfit}.
As the NSCs are more extended than the PSFs, no explanation for the diverging S{\'{e}}rsic index could be determined.
To quantify the extent of the affected NSCs, we fixed the S{\'{e}}rsic index to a value of $n = 2$.
The choice of this value was motivated by the recent work of \citet{pechetti2020a} who classified their fits into three categories.
NSCs which could be fit `well' (their `Quality 0' fits) have a mean / median value of $n = 1.9 \, / \, 2.9$.
Although six out of their 17 NSCs have $n > 3$, we decided to set $n = 2$ and to determine a systematic uncertainty based on fits using $n = 0.5$ and $n = 3$.
In the parameter range $n \in [0.5, 3.0]$, the S{\'{e}}rsic index does not correlate with the effective radius, allowing us to put constraints on it.
For larger S{\'{e}}rsic indices, the effective radius also increases in a non-linear way.
We give more details and discuss this choice further in \Cref{subsubsec:fixation_of_sersic_indices}.
However, it will become evident in \Cref{sec:results_and_discussion} that the key results of this paper remain unchanged.
\begin{table}
  \small
  \caption{%
    Parameters and their boundaries supplemented to {\imfit}.
    The same values are used for all galaxies and filters.
  }
  \begin{center}
    \begin{threeparttable}
      \begin{tabular}{%
        l
        l
        l
        l
      }
        \toprule
        \multicolumn{1}{c}{Parameter} & \multicolumn{1}{c}{Boundary} & \multicolumn{1}{c}{Unit} & \multicolumn{1}{c}{Description}\\
        \midrule

        $x_{0}$            & $[45, \, 55]$                             & \si{\pixel}  & NSC position \\
        $y_{0}$            & $[45, \, 55]$                             & \si{\pixel}  & NSC position \\
        PA                 & $[-359.99, \, 359.99]$\tnote{(a)}         & \si{\deg}    & Position angle \\
        $\epsilon$         & $[0.00, \, 0.99]$                         & {-{-}}       & Ellipticity \\
        $n$                & $[0.00, \, 15.00]$                        & {-{-}}       & S{\'{e}}rsic index \\
        $r_{\mathrm{eff}}$ & $[0.00, \, 50.00]$                        & \si{\pixel}  & Effective radius \\
        $I_{\mathrm{eff}}$ & $[0.00, \, I_{\mathrm{max}}]$\tnote{(b)}  & counts       & Intensity at $r_{\mathrm{eff}}$ \\

        \bottomrule
      \end{tabular}
      \begin{tablenotes}
        \item[(a)] Often the fit was stuck at a boundary of \SI{0}{\degree}, hence the extension towards negative values. If the best fit position angle was negative, we added \SI{180}{\degree} (twice) until it became positive.
        \item[(b)] $I_{\mathrm{max}}$ is the peak intensity of the nuclear star cluster.
  \end{tablenotes}
    \end{threeparttable}
  \end{center}
  \label{tab:parameter_boundaries}
\end{table}

\subsection{Nuclear star cluster stellar mass}
\label{subsec:nuclear_star_cluster_stellar_mass}

Integrating \Cref{equ:sersic_profile} over the radial component while assuming an ellipticity ($\epsilon$) yields the total intensity of the NSC ($L$) as
\begin{equation}
  \label{equ:nsc_intensity}
  L = 2 \pi (1 - \epsilon) r_{\mathrm{eff}}^{2} I_{\mathrm{eff}} \times \frac{n \, \mathrm{e}^{b_{n}}}{(b_{n})^{2n}} \times \Gamma(2n) \quad .
\end{equation}
Combining $L$ with the zeropoint magnitudes and exposure times, which are both given in \Cref{tab:available_data}, allows the calculation of apparent magnitudes.

We derived stellar masses using the $V-I$ colour and, therefore, converted from {\hst} magnitudes to the $BVRI$ system.
Following the approach by \citet{pechetti2020a}, magnitudes were converted using different synthetic transformation.
For the ACS/WFC data, the magnitudes were transformed using Table~22 and Equation~12 of \citet{sirianni2005a}.
WFPC2/WF and WFPC2/PC magnitudes were converted using Table~4 and Equation~16 of \citet{dolphin2009a}.

Once the magnitudes were transformed, we corrected them for Galactic extinction using a recalibrated version of the \citet{schlegel1998a} dust maps \citep{schlafly2011a} and assuming the reddening law of \citet{fitzpatrick1999a} with $R_{V} = 3.1$.
The corrected apparent magnitudes were then used to determine absolute magnitudes via the galaxy distance estimates and the absolute magnitude of the Sun\footnote{Obtained from \url{http://mips.as.arizona.edu/~cnaw/sun.html}.}.
All extinction corrected apparent magnitudes are presented in \Cref{tab:nsc_parameters}.

The stellar mass-to-light ratio relies on the $I$-band luminosity and $(V - I)_{0}$ colour and is identical to the one used in \citet{pechetti2020a}.
This relation ($M_{\star} / L_{I}$) is based on the work of \citet{roediger2015a} and reads
\begin{equation}
  \label{equ:mass_to_light_ratio}
  \log_{10} M_{\star} / L_{I} = -0.694 + 1.335 \times (V - I)_{0} \quad ,
\end{equation}
where the slope and intercept have been determined by fitting a linear relationship to the underlying data which was provided by Joel Roediger (private communication).

The uncertainty on the NSC stellar masses are dominated by the uncertainty on the mass-to-light ratio which we assume to be \SI{0.3}{\dex} \citep{roediger2015a}.
Other uncertainties, which have been included via Gaussian error propagation, include the statistical and systematic uncertainties of the fit (see \Cref{sec:assessing_uncertainties}), the uncertainty on the absolute magnitude of the Sun (assumed to be \SI{0.04}{\mag}), and the uncertainty on the distance estimates.
All quoted uncertainties give the $1\sigma$ interval.
The resulting parameter values and their uncertainties are presented in \Cref{tab:nsc_parameters}.

\section{Results}
\label{sec:results_and_discussion}

In total, we derive NSC structural parameters for \num{19} objects. 
In the case of dw{\,}1335-29, the signal-to-noise ratio of the ACS/SBC \textit{F150LP} data were to low to allow for an accurate determination of NSC parameters.
In the case of PGC{\,}154449, we could not determine parameter estimates from either the ACS WFC \textit{F606W} or \textit{F814W} data as the effective radius was approaching the boundary of \SI{50}{\pixels} in all attempts.
We changed the size of the fitting region, the fitting routine, and applied various masks without achieving a stable fit result.
For UGC{\,}01104, structural parameters could not be determined in the ACS/WFC \textit{F300W} band.

Furthermore, we derive colours and stellar mass estimates for \num{17} objects. 
The blue colour estimate of ESO{\,}553-046 [$(V - I)_{0} \sim \SI{-3.2}{\mag}$, \textit{cf}. \Cref{tab:nsc_parameters}] leads to an unreliable estimate of the NSC mass.
As no structural parameters could be estimated in two filters for dw{\,}1335-29, PGC{\,}154449, and UGC{\,}01104, we do not derive NSC stellar masses.
Finally, the stellar mass-to-light ratio of four NSCs is unreasonably high ($M_{\star} / L_{I} \gtrsim 4 \mathrm{M}_{\odot} / \mathrm{L}_{\odot}$).
These data points are excluded from the surface brightness and mass density profiles (\textit{cf}.~\Cref{subsubsec:surface_brightness}) and the determination of the scaling relation between NSC and host galaxy stellar mass (\textit{cf}.~\Cref{subsec:nsc_stellar_mass_versus_galaxy_stellar_mass}).

\subsection{Literature data}
\label{subsec:literature_data}

We compare our results to other NSCs in the Local Volume, in massive late-type field galaxies, and in dwarf ellipticals in the core of the Virgo cluster.
For the Local Volume, we selected all known nucleated galaxies and obtained their NSC parameters, where available, from the most recent literature reference identified by \citet[][their Table~D1]{hoyer2021a}.
For NSCs in massive late-type field galaxies, we used the data tables of \citet{georgiev2014a}.
As the authors do not provide stellar masses, and to avoid systematic differences to our approach, we adopted their \textit{F606W} and \textit{F814W} apparent magnitudes and repeated the steps outlined in \Cref{subsec:nuclear_star_cluster_stellar_mass}.
Table~5 of \citet{sanchez-janssen2019a} provides stellar masses for NSCs in dwarf ellipticals in the core of the Virgo cluster.
In addition, we adopt the data from \citet{carlsten2022a} for dwarfs around massive late-type field galaxies.
We compare to Galactic globular clusters using the data from \citet{harris1996a} and \citet{baumgardt2018b}.

We present an overview of the parameters of other NSCs in Local Volume galaxies in \Cref{tab:lit_nsc_parameters}.
NSC stellar masses for the sample of \citet{georgiev2014a} are presented in \Cref{tab:lit_georgiev2014}.

Galaxy stellar masses were adopted from \citet{hoyer2021a} for the whole Local Volume data set and the galaxy sample of \citet{georgiev2014a}.
We take galaxy stellar masses for dwarf ellipticals in the core of the Virgo cluster from Table~4 of \citet{sanchez-janssen2019a}.

\subsection{Wavelength dependence}
\label{subsubsec:wavelength_dependence}

We investigate whether NSC structural parameters are wavelength dependent by comparing differences in parameter estimates between the most commonly available \textit{F660W} and \textit{F814W} bands.
Within the uncertainties, we find no significant differences in both $\epsilon$ and $r_{\mathrm{eff}}$.
The position angle changes insignificantly ($\Delta \mathrm{PA} \lesssim \SI{30}{\degree}$) for most NSCs.

\subsection{Structural properties}
\label{subsec:structural_properties}

Here we investigate the structural properties of the new detections using the \textit{F814W} band.
We compare to other data from the Local Volume and \citet{georgiev2014a} using the same band, if available.%
\footnote{For the Local Volume data set, we use the reddest band in case the \textit{F814W} is unavailable.}
In addition, we compare to the globular cluster population of the Milky Way \citep{baumgardt2018b,harris1996a}.

\begin{figure*}
  \centering
  \includegraphics[width=\textwidth]{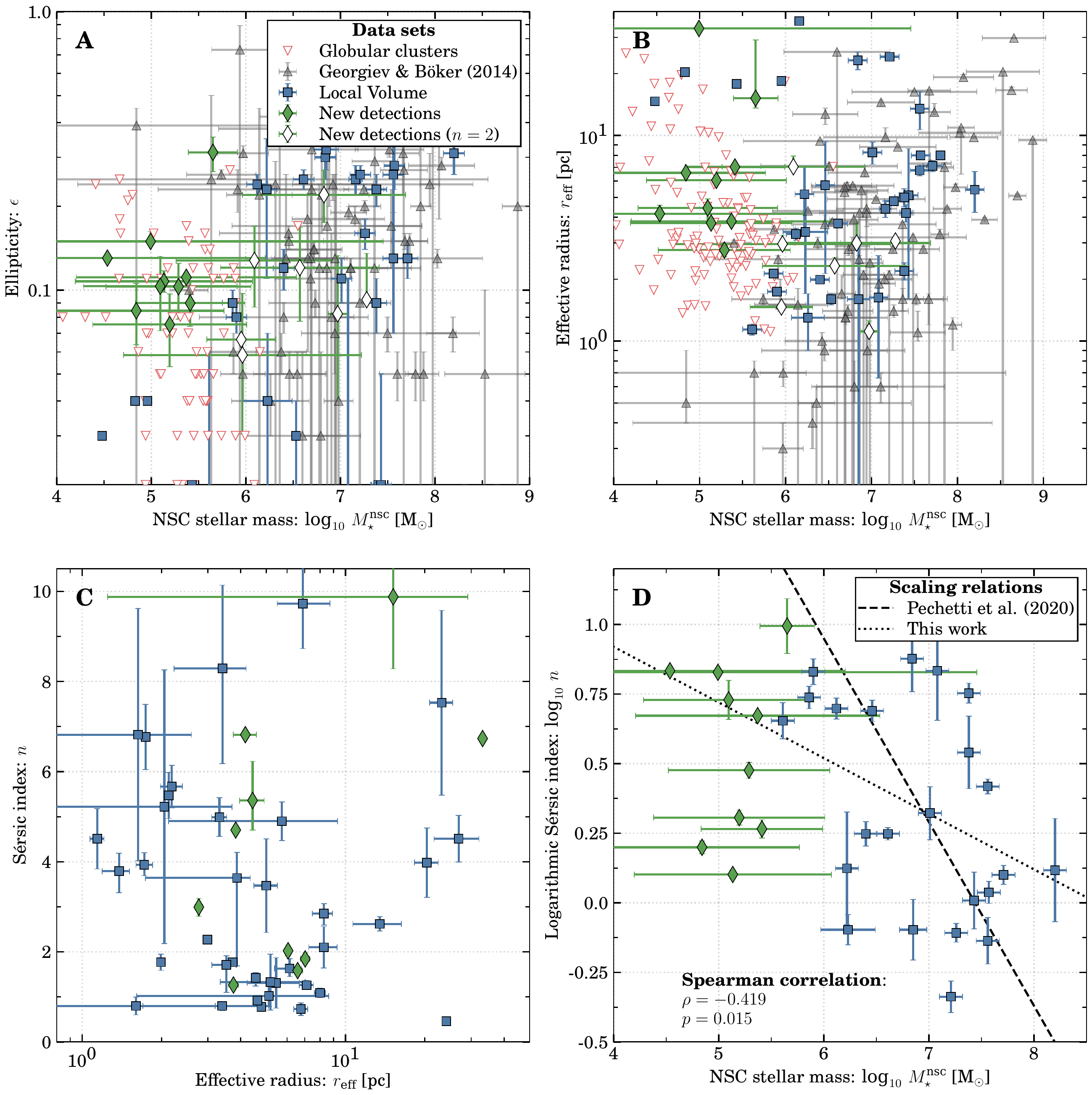}
  \caption{%
    \textit{Panel A}:
    Ellipticity ($\epsilon$) versus nuclear star cluster (NSC) stellar mass ($M_{\star}^{\mathrm{nsc}}$).
    We compare the new detections (green diamonds) to NSCs in massive late-type spirals \citep{georgiev2014a}, a compilation of NSCs in the Local Volume, and Galactic globular clusters \citep{harris1996a,baumgardt2018b}.
    The new detections are split into two categories, depending on whether the S{\'{e}}rsic index ($n$) needed to be fixed at $n = 2$.
    \textit{Panel B}:
    Effective radius ($r_{\mathrm{eff}}$) versus NSC stellar mass.
    The markers and color are the same as in panel A.
    \textit{Panel C}:
    S{\'{e}}rsic index versus effective radius.
    The data from \citet{georgiev2014a} and \citet{harris1996a,baumgardt2018b} are not available as the clusters were modelled with King profiles.
    Most of the Local Volume data come from \citet{pechetti2020a}.
    \textit{Panel D}:
    S{\'{e}}rsic index versus NSC stellar mass.
    A dashed black line gives the weak scaling relation identified by \citet{pechetti2020a}.
    The dotted line gives the best-fit linear relation including the new detections.
    The Spearman correlation index $\rho$ and its associated $p$-value of thew new fit are given in the lower left corner.
  }
  \label{fig:structure}
\end{figure*}

Panel A of \Cref{fig:structure} shows the ellipticity versus NSC stellar mass.
Most of the new detections have $\epsilon \sim \num{0.1}$ but at most $\sim \num{0.3}$.
With the exception of the most massive NSCs, both the stellar mass and ellipticity compare to Milky Way GCs.
The overall increase of ellipticity with increasing mass is in agreement with Figure~24 of \citet{spengler2017a}.
The new detections reveal that this trend does not continue down to the lowest mass clusters, as suggested by the few existing Local Volume data points from the literature.
Similarly, the GC population of the Milky Way does not show a correlation as well.

Panel B shows the effective radius versus NSC mass.
The new detections occupy the low-mass and compact-size region in the parameter space.
While at higher NSC mass there exists a correlation between the effective radius and NSC mass \citep[e.g.][]{georgiev2014a,georgiev2016a,neumayer2020a}, this relation appears to break down at $M_{\star}^{\mathrm{nsc}} \sim \SI{e6}{\Msun}$, as revealed by the new detections.
The distribution of Galactic GCs overlap with the new detections, corroborating a tight connection between both types of systems in this mass range.
Furthermore, the new detections appear to follow the same trend as the GCs by increasing in effective radius with decreasing mass.

There exist six data points with $r_{\mathrm{eff}} \geq \SI{10}{\pc}$ and $M_{\star}^{\mathrm{nsc}} \leq \SI{e6}{\Msun}$, which partially overlap with the distribution of Galactic GCs but are otherwise outliers from the NSC distribution.
If the NSCs truly reside in this part of the parameter space, one explanation could be that their evolution is similar to that of the NSC of the Pegasus dwarf galaxy:
the cluster initially formed in the centre of their host galaxy, was relocated outside of the central region where $r_{\mathrm{eff}}$ increased due to the weaker tidal field, and migrated back towards the centre \citep{leaman2020a}.
For both UGC{\,}08638 and NGC{\,}4163, this mechanism could still be in process as the projected distance of the NSC to the photometric centre is \SI{\sim 480}{\pc} and \SI{\sim 150}{\pc}, respectively \citep{georgiev2009a}.
The projected distance of the other two galaxies (KK{\,}197 and ESO{\,}269-066) is close to \SI{0}{\pc} \citep{georgiev2009a}.

Panel C of \Cref{fig:structure} shows the S{\'{e}}rsic index versus effective radius.
There appears to be a trend in that the index drops from $n \sim 7$ to $\sim 1$ when the effective radius increases from $r_{\mathrm{eff}} \sim \SI{1}{\pc}$ to $\sim \SI{10}{\parsec}$.
However, multiple NSCs occupy the high S{\'{e}}rsic index and high effective radius parameter space, questioning a potential universal correlation.
More data and further studies are required to explore this parameter space.

Panel D shows the previously identified weak relationship between the logarithmic S{\'{e}}rsic index and NSC stellar mass by \citet{pechetti2020a}.
We add the new detections to the figure and fit the combined data sets with a linear function.
The best fitting relation reads
\begin{equation}
  \label{equ:n_mass}
    \log_{10} \, n_{F814W} = 0.52^{+0.50}_{-0.38} - 0.20^{+0.53}_{-0.29} \times \log_{10} \frac{M_{\star}^{\mathrm{nsc}}}{\SI{e6}{\Msun}} \; .
\end{equation}
The parameters differ significantly from the values found by \citet{pechetti2020a} and question the presence of a tight correlation.
Therefore, while the Spearman correlation coefficient evaluates the trend as significant ($p = 0.015$), we recommend against using the S{\'{e}}rsic index relation to parameterise NSCs.

\subsection{Surface brightness \& surface mass density profiles}
\label{subsubsec:surface_brightness}

\begin{figure}
  \centering
  \includegraphics[width=\columnwidth]{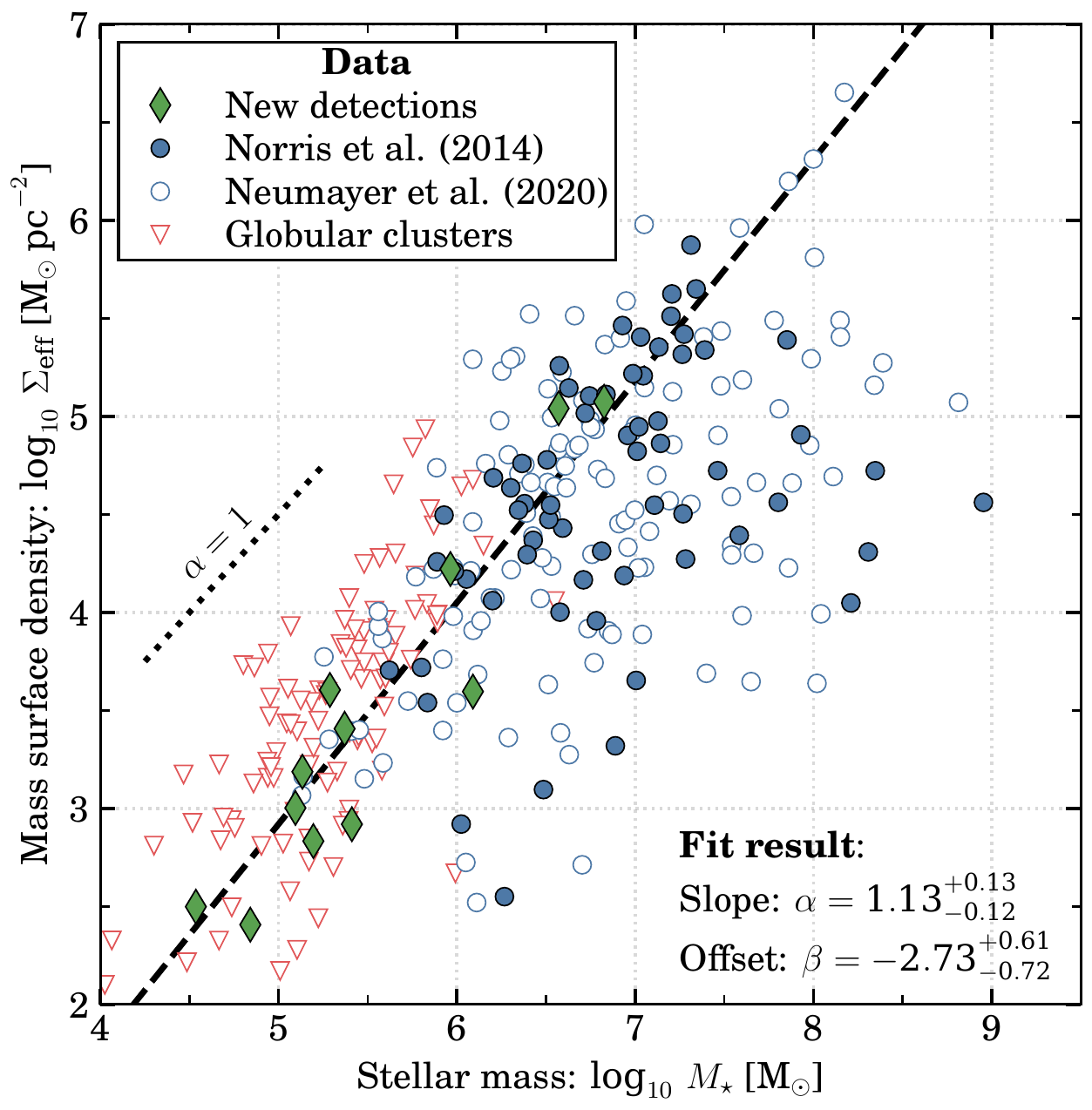}
  \caption{%
    Mean stellar mass surface density within the effective radius ($\Sigma_{\mathrm{eff}}$) versus cluster mass ($M_{\star}$).
    We compare the new detections (green diamonds) to other nuclear star clusters from \citet{norris2014a} and \citet{neumayer2020a}, and to Milky Way globular clusters \citep{baumgardt2018b}.
    The best-fit values of a linear relationship fit to the new detections, as determined through \num{e5} bootstrap iterations, are indicated in the lower right corner.
  }
  \label{fig:density_mass}
\end{figure}

Combining the effective radius and stellar mass, we determine a mean surface density for the new detections.
We show this parameter space in \Cref{fig:density_mass}, comparing the new detections with literature data from \citet{norris2014a} and \citet{neumayer2020a} for other NSCs, and with \citet{baumgardt2018b} for Milky Way GCs.
For the newly detected NSCs, we fit the correlation using a linear function to find
\begin{equation}
  \label{equ:sigma_mass}
  \log_{10} \, \Sigma_{\mathrm{eff}} = -2.72^{+0.61}_{-0.71} + 1.13^{+0.13}_{-0.12} \times \log_{10} \, M_{\star}^{\mathrm{nsc}} \; ,
\end{equation}
where the parameter values are determined through \num{e5} bootstrap iterations.
We note that although some of NSCs at $M_{\star}^{\mathrm{nsc}} \sim \SI{e8}{\Msun}$ seem to follow this relation as well, their overall distribution get wider and seems to flatten.
At the low-mass end, the newly detected NSCs overlap again with Galactic GCs.
Note that about \SI{65}{\percent} of these GCs fall above the best-fit relationship.

\begin{figure*}
  \centering
  \includegraphics[width=\textwidth]{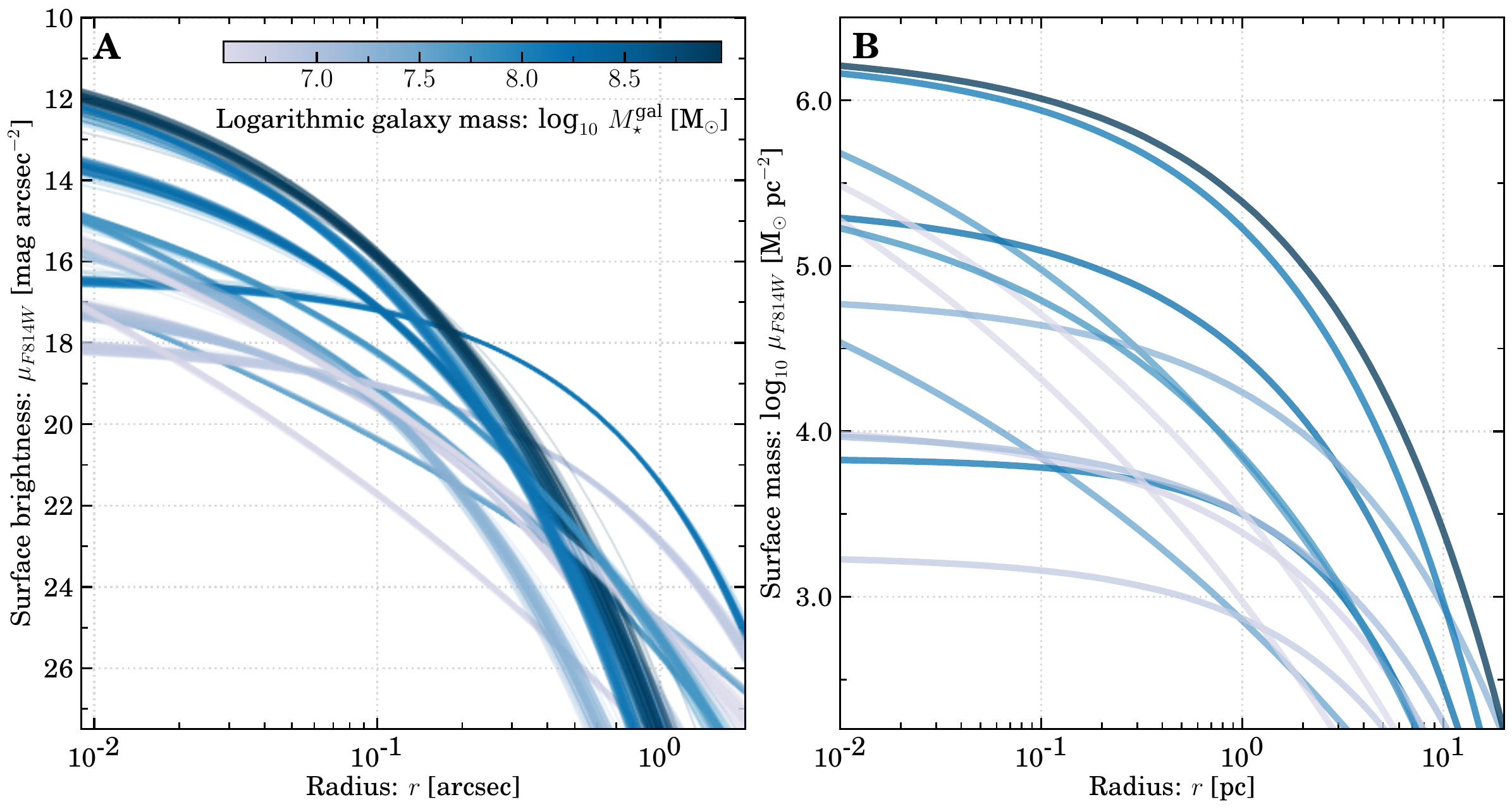}
  \caption{%
    \textit{Panel A}:
    Surface brightness in the \textit{F814W} band ($\mu_{\mathrm{\textit{F814W}}}$) versus radius ($r$) of the newly detected nuclear star clusters (NSCs).
    The profiles give the S{\'{e}}rsic models fit to the data.
    To highlight uncertainties, we show \num{100} profiles for each NSC, randomly drawn from a total of \num{500} bootstrap iterations.
    Each set of profiles is colour-coded by the stellar mass of the host galaxy where a darker colour corresponds to a more massive galaxy.
    \textit{Panel B}:
    Surface mass density based on the \textit{F814W} band versus radius.
    The conversion from the best-fit surface brightness profiles to mass profiles assumes a radially constant mass-to-light ratio.
    This assumption is invalid in higher-mass NSCs \citep{carson2015a,feldmeier-krause2015a,feldmeier-krause2017a}, which is why we show only a single profile for each NSC.
    The colour of the lines is the same as in panel A.
  }
  \label{fig:sb}
\end{figure*}

Next, we explore the surface brightness of the star clusters.
Panel A of \Cref{fig:sb} shows the surface brightness as a function of radius where the profiles relate to the S{\'{e}}rsic model fits from the \textit{F814W} band.
To highlight uncertainties, we plot the profiles of \num{100} out of \num{500} bootstrap iterations, which we used to determine statistical uncertainties (\textit{cf}.~\Cref{subsubsec:statistical_uncertainties}).
Each set of profiles is colour-coded based on the host galaxy stellar mass.

Similarly, panel B shows the surface mass profile versus radius.
We convert the profiles to stellar mass by using the mean colour of the NSCs.
Note the assumption that the mass-to-light ratio is radially constant, which is not the case for higher mass NSCs in other Local Volume galaxies \citep[e.g.][]{carson2015a} and the Milky Way NSC \citep{feldmeier-krause2015a,feldmeier-krause2017a}.
However, as a function of wavelength, the size or ellipticity does not differ significantly for the new detections and, therefore, using the mean colour likely provides a decent estimate of their average stellar populations.
The colour-coding of the profiles is the same as in panel A.

\begin{figure*}
  \centering
  \includegraphics[width=\textwidth]{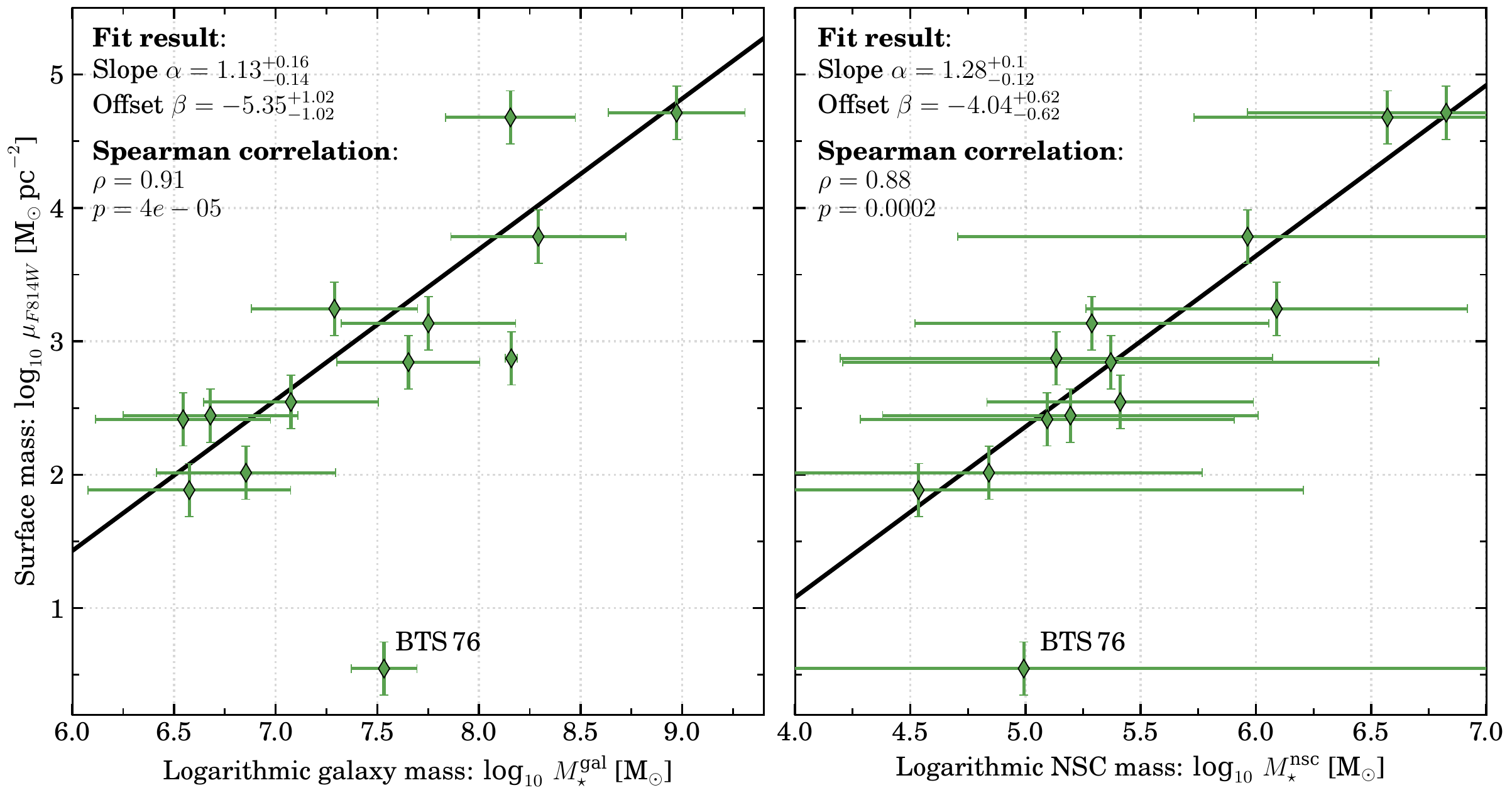}
  \caption{%
    Surface mass profile evaluated at the nuclear star clusters effective radius versus host galaxy (\textit{panel A}) and cluster stellar mass (\textit{panel B}).
    Solid lines give the best-fit linear relation ($\log_{10} \, \mu = \alpha + \beta \times \log_{10} \, M_{\star}$) whose parameters, as determined through \num{e5} bootstrap iterations, are indicated in the panels.
    In addition, we show the Spearman correlation parameter ($\rho$) and its associated $p$-value.
    BTS{\,}76, as indicated, does not fit the overall trend and was excluded from the fits.
  }
  \label{fig:mueff_mass}
\end{figure*}

From both figures it becomes apparent that the surface brightness of the clusters positively correlates with the host galaxy stellar mass.
To quantify this observation further, we show the surface mass profiles evaluated at the clusters' effective radii versus the host galaxy and NSC stellar masses in \Cref{fig:mueff_mass}.
According to the Spearman correlation coefficients, we find a clear correlation between both quantities.
A fit using a linear relationship yields
\begin{equation}
  \label{equ:mu_eff_masses}
  \log_{10} \, \mu = 
  \begin{cases}
    1.13^{+0.16}_{-0.14} - 5.3^{+1.0}_{-1.0} \times \log_{10} \, M, \quad \mathrm{for} \; M = M_{\star}^{\mathrm{gal}} \\
    1.29^{+0.10}_{-0.12} - 4.05^{+0.62}_{-0.62} \times \log_{10} \, M, \quad \mathrm{for} \; M = M_{\star}^{\mathrm{nsc}} \;,
  \end{cases}
\end{equation}
where the uncertainties are determined with \num{e5} bootstrap iterations.
Note that the slope value for the relation using the NSC mass is steeper than one.
This is related to both the NSC versus host galaxy stellar mass relation (\textit{cf}.~\Cref{subsec:nsc_stellar_mass_versus_galaxy_stellar_mass} below) and the observation that the effective radius decreases with increasing NSC mass for the new detections (\textit{cf}.~\Cref{fig:structure}, panel B).

From both panels it is apparent that BTS{\,}76 does not follow the relationship and was excluded from both fits.
Compared to other NSCs, the effective radius of this nucleus is significantly larger.
As discussed in \Cref{subsec:structural_properties}, the NSC sub sample with large effective radii and low stellar masses may have evolved differently from the other clusters:
if the cluster relaxes in a weaker tidal field (i.e.\ the outskirts of the host galaxy), its central density may drop while the total mass of the cluster remains roughly the same.

Note that \citet{pechetti2020a} investigated the three-dimensional density of high-mass NSCs in higher-mass galaxies finding a similar trend:
the NSC density positively scales with the host galaxy stellar mass.
Our data show that such a correlation appears to continue down to lower galaxy and NSC stellar masses, effectively extending the existence of a relation from $\log_{10} \, M_{\star}^{\mathrm{gal}} / \mathrm{M}_{\odot} \sim 11$ to $\log_{10} \, M_{\star}^{\mathrm{gal}} / \mathrm{M}_{\odot} \sim 6.5$.

\subsection{NSC stellar mass versus galaxy stellar mass}
\label{subsec:nsc_stellar_mass_versus_galaxy_stellar_mass}

In this section we investigate the scaling relation between the NSC stellar mass and its host stellar mass.
We combine literature data of the Local Volume, \citet{georgiev2014a}, and \citet{carlsten2022a} with our new detections to gain statistical significance.
To this combined data set, we fit the function 
\begin{equation}
  \label{equ:mass_relation}
  \log_{10} \, M_{\star}^{\mathrm{nsc}} = \alpha \times \log_{10} \, \frac{M_{\star}^{\mathrm{gal}}}{\SI{e9}{\Msun}} + \beta \quad ,
\end{equation}
which has also been used previously \citep[][]{georgiev2016a,neumayer2020a}.
To fit the data, we use the \textsc{scipy} implementation of the orthogonal distance regression, which takes into account uncertainties on both axes (see also \citealp{boggs1990a}).
The uncertainty on the stellar masses of literature data are assumed to be \SI{0.3}{\dex} if no value is provided.
As the slope $\alpha$ of the relation in \Cref{equ:mass_relation} seems to steepen for galaxies above $M_{\star}^{\mathrm{gal}} \sim \SI{e9.5}{\Msun}$ \citep{georgiev2016a,neumayer2020a}, we restrict the fit to $M_{\star}^{\mathrm{gal}} < \SI{e9.5}{\Msun}$.
Furthermore, from the fit we removed four NSCs (DDO{\,}133, LV{\,}J1205+2813, NGC{\,}5011C, and UGC{\,}04998) as they have high stellar mass-to-light ratios ($M_{\star} / L_{I} \gtrsim 4 \mathrm{M}_{\odot} / L_{\odot}$).
The final uncertainties of the fit were determined via \num{e5} bootstrap iterations.

\begin{figure*}
  \centering
  \includegraphics[width=0.7\textwidth]{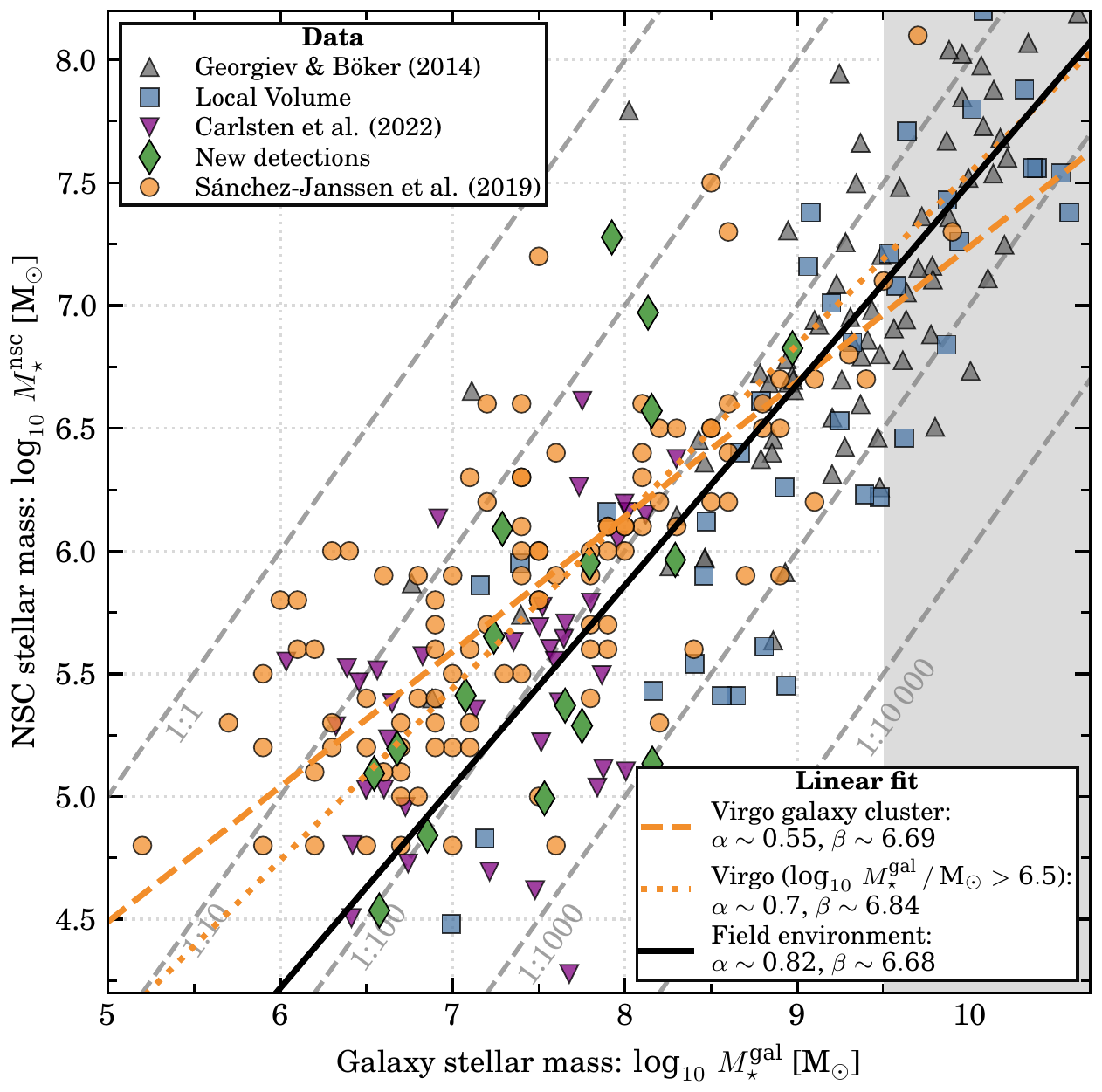}
  \caption{%
    Nuclear star cluster (NSC) stellar mass ($M_{\star}^{\mathrm{nsc}}$) versus host galaxy stellar mass ($M_{\star}^{\mathrm{gal}}$) for the new detections (greend diamonds), a compilation of Local Volume data (blue squares), massive late-type galaxies in the field (\citealp{georgiev2014a}; gray up-pointing triangles), dwarf galaxies around massive late-types (\citealp{carlsten2022a}; purple down-pointing triangles), and dwarfs in the core of the Virgo galaxy cluster (\citealp{sanchez-janssen2019a}; orange circles).
    Uncertainties are omitted for clarity.
    The combined data of new detections, other Local Volume, and field galaxies are fit with a linear relationship, such that $\log_{10} \, M_{\star}^{\mathrm{nsc}} = \alpha \log_{10} \, (M_{\star}^{\mathrm{gal}} \, / \, \SI{e9}{\Msun}) + \beta$.
    This relationship is shown with a black solid line; its best-fit parameters are indicated in the bottom right.
    A gray shaded area gives the region excluded from the fit.
    Furthermore, from the fit we exclude four NSCs (DDO{\,}133, LV{\,}J1205+2813, NGC{\,}5011C, and UGC{\,}04998) which have a high stellar mass-to-light ratio, but still show the data points in the figure.
    Uncertainties are determined thorugh \num{e5} bootstrap iterations.
    For dwarfs in the core of the Virgo cluster, we perform the same fit and show the best-fit relation with a dashed orange line.
    The slope of $\alpha \sim 0.55$ is significantly below the value found for dwarfs in the field.
    Restricting the fit to galaxies in Virgo to $M_{\star}^{\mathrm{gal}} \geq \SI{e6.5}{\Msun}$ (orange dotted) line results in a slope with agrees with the one for the field environment within the $1 \sigma$ interval.
  }
  \label{fig:nscmass_galmass}
\end{figure*}

\Cref{fig:nscmass_galmass} shows the data set as well as the best-fit relationship for which we find $\alpha = 0.82^{+0.08}_{-0.08}$ and $\beta = 6.68^{+0.13}_{-0.13}$.
This slope is steeper than what was found by \citet[][$\alpha \sim 0.48$]{neumayer2020a} who used data from various publications and a mix of environments.
Restricting the fit to the high-mass end yielded a value of $\alpha \sim 0.92$, which agrees with a previously reported value \citep{georgiev2016a} and our value.

Our results and the observation that the fit by \citet{neumayer2020a} is dominated by dwarfs in a dense cluster environment [Virgo \citep{sanchez-janssen2019a} and Fornax \citep{ordenes-briceno2018b}] could suggest that the environment of the dwarf galaxies plays a role in the NSC versus host stellar mass relationship.
To test this hypothesis, we add the data set of \citet{sanchez-janssen2019a}, exploring the relationship for dwarfs in the core of the Virgo galaxy cluster.
Only considering their data, we find $\alpha = 0.55^{+0.06}_{-0.05}$ and $\beta = 6.69^{+0.10}_{-0.09}$ using again \num{e5} bootstrap iterations.
As expected, the slope is comparable to the value found by \citet{neumayer2020a} but significantly smaller than the value for dwarfs in the field.

To check whether the origin for the difference between environments stems from the lowest mass galaxies, we repeat the fit to the Virgo cluster data set forcing $M_{\star}^{\mathrm{gal}} \geq \SI{e6.5}{\Msun}$.
This results in $\alpha = 0.70^{+0.08}_{-0.07}$ and $\beta = 6.83^{+0.13}_{-0.11}$.
As the slope is now comparable to the one found for the field environment we conclude that the low-mass galaxies in the Virgo cluster, which host more massive NSCs than in the field, are responsible for environmental trends.

\section{Discussion}
\label{sec:discussion}

  We presented a comparison between the Milky Way GCs and the newly detected NSCs in the previous sections and argue in \Cref{subsec:formation_scenario} that dissipationless GC migration is the main formation scenario for NSCs in low-mass dwarf galaxies.
  Afterwards, in \Cref{subsec:are_the_nscs_a_merger_product_of_gcs}, we discuss whether the NSCs are a merger product of multiple GCs or whether they are not.

\subsection{Formation scenario}
\label{subsec:formation_scenario}

NSCs are believed to form via two mechanisms:
at the low-mass end, GC migration appears to dominate the formation of NSCs \citep[e.g.][]{tremaine1975a,hartmann2011a,antonini2015b,fahrion2022b} and \textit{in-situ} star formation contributes only a small part to the mass budget, if at all.
With increasing galaxy stellar mass \textit{in-situ} star formation gains importance \citep[e.g.][]{turner2012a,sanchez-janssen2019a,neumayer2020a} and will eventually dominate over the GC migration scenario \citep{fahrion2021a,fahrion2022a}.

We compared the structural properties of the newly detected NSCs with Milky Way GCs in \Cref{fig:structure,fig:density_mass} finding a similarity between both systems.
More specifically, the ellipticity, effective radius, stellar mass, and surface density of many of the new detections matches the distribution of Milky Way GCs.
As speculated in the literature already \citep[e.g.][]{miller2007a,sanchez-janssen2019a}, this is a direct hint that the dissipationless GC migration scenario is the main formation mechanism of these NSC.

\begin{figure}
  \centering
  \includegraphics[width=\columnwidth]{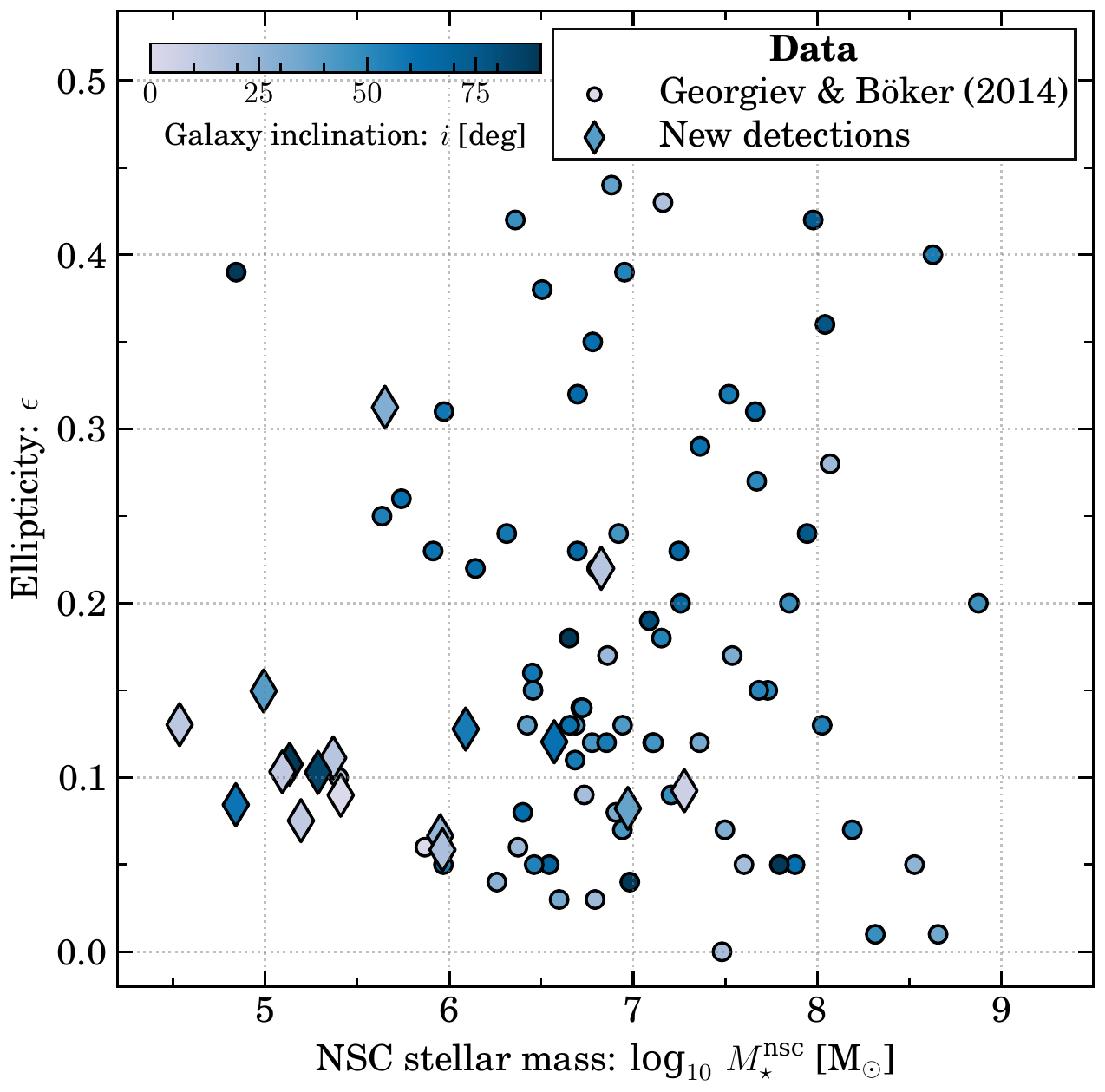}
  \caption{%
    Ellipticity ($\epsilon$) versus nuclear star cluster stellar mass ($M_{\star}^{\mathrm{nsc}}$).
    We show the new detections (diamonds) and compare with NSCs in massive late-type galaxies \citep{georgiev2014a}.
    Each data point is colour-coded by the inclination of the host galaxy.
  }
  \label{fig:inclination}
\end{figure}

We also found that the ellipticity remains roughly constant below $M_{\star}^{\mathrm{nsc}} \sim \SI{e6.5}{\Msun}$ and starts to increase for higher mass clusters.
An increase in ellipticity hints towards \textit{in-situ} star formation as the in-falling gas is expected to form stars in a flattened disk due to its angular momentum.
In observations, such a flattening has been observed in combination with young stellar populations in edge-on spiral galaxies \citep[e.g.][]{seth2006a}.
In simulations, \citet{hartmann2011a} showed that NSCs, which formed through repeated GC mergers, typically are not very flattened.
Crucially, as we show in \Cref{fig:inclination} the measured ellipticity of the NSCs does not depend on the inclination of the host galaxy at all stellar masses.

As shown in \Cref{fig:nscmass_galmass}, the NSC versus host galaxy stellar mass correlation appears to be affected by host environment with cluster members typically hosting more massive NSCs than field galaxies.
We found that the difference is greatest at the low-mass end $M_{\star}^{\mathrm{gal}} \leq \SI{e6.5}{\Msun}$ and becomes insignificant towards higher masses.
If \textit{in-situ} star formation is unimportant at the lowest galaxy stellar masses, the difference in NSC must arise from differences in the progenitor GCs.

There appear two possibilities to generate more massive NSCs:
\begin{itemize}
  \item[1.]
  The NSCs in dwarfs in dense environments experienced more GC merger events than NSCs in a field environment, elevating their masses.
  We discuss this option further in \Cref{subsec:are_the_nscs_a_merger_product_of_gcs}.

  \item[2.]
  The difference in mass does not arise from a significant difference in mergers but from a difference in progenitor GC mass.
\end{itemize}
The argument that the progenitor GC is more massive in a dense environment relies on GC formation scenarios.
The cluster formation efficiency \citep{bastian2008b} positively correlates with the surface density of star formation \citep[see][and references therein]{stahler2018a} and leads to an elevated mass fraction of stars in clusters.
From observations it appears that this effect results in an increased number of GCs in present-day dwarf galaxies in galaxy clusters \citep[e.g.][]{peng2008a} and not in differences in the GC mass function \citep{carlsten2022a}.
The GC luminosity function appears to be roughly equivalent between the environments \citep[Figure~7 in ][]{carlsten2022a} but this may not be the case at high redshift \citep[e.g.][]{parmentier2007a,kruijssen2012b}.

If the GC mass function remains unchanged between environments at the time when the NSC formed, the NSC's mass may still be elevated due to the higher number of GCs produced in a dense environment.
When drawn from the same distribution, a higher number of GCs correspond to a higher probability that the most massive GC in a galaxy in a dense environment is more massive than its counterpart in a galaxy in a loose environment.

We note that the differences in NSC stellar mass found at the low-mass end could also be related to selection bias.
Our data rely on a catalogue of galaxies in the Local Volume \citep[see][and references therein]{karachentsev2013a} while the Virgo cluster data of \citet{sanchez-janssen2019a} relies on a uniform set of imaging data \citep{ferrarese2012a}.
The data of \citet{carlsten2022a} indicate that satellites around massive field galaxies, where a significant mass fraction is contained by the NSC, do exist but in fewer numbers than in the Virgo cluster.
Whether this is also a selection effect is unclear.
Note that it appears unlikely that higher-mass galaxies were stripped by $\sim \SI{1}{dex}$ in mass in the galaxy cluster while the NSC mass remains unchanged \citep[e.g.][]{smith2016c}.

Truncated star formation during the galaxy infall may lead to a bias in the NSC versus galaxy mass relationship as well:
asynchronous formation timescales of the NSC and its host galaxy leads to a higher cluster mass fraction if most cold gas is removed during infall.
This effect could partly be responsible for both the observed environmental dependence of the stellar mass correlation as well as a higher NSC occupation fraction in dense environments \citep{leaman2022a}.
Whether this effect can fully explain the observed environmental dependence remains unclear.

\subsection{Are our newly detected NSCs merger products of GCs?}
\label{subsec:are_the_nscs_a_merger_product_of_gcs}

A second method for forming NSCs is the process of repeated GC mergers.
As mentioned in the previous section, at fixed galaxy stellar mass, the number of GCs is higher in a dense environment than in the field.
Therefore, a present-day NSC in a galaxy in a dense environment could have experienced more GC mergers than in a loose environment, explaining its increased mass at the low-mass end of galaxies.

One argument in favor of this scenario is shown in \Cref{fig:density_mass}.
We found that $\sim \SI{65}{\percent}$ of GCs fall above the mass density versus cluster mass relation.
\cite{antonini2012b} and \cite{antonini2013a} showed that the merger product of two GCs results in an increase in effective radius of the merger product where $r_{\mathrm{eff}} \propto \sqrt{M_{\star}}$.
If two clusters merge in the density versus mass parameter space, their mass will increase but the overall density will drop, meaning that the data point moves towards the bottom right part in \Cref{fig:density_mass}.
Therefore, as the Milky Way GCs are, on average, denser at the same stellar mass than our new detections, NSCs could be a merger product of multiple progenitor GCs.
However, given the uncertainties of the data points it is not possible to prove this scenario for individual objects.

If true in all environments, we would expect that the effective radius of NSCs in the core of the Virgo cluster are more extended than in the field environment, as they experienced more GC merger events.
The data of the Next Generation Virgo Cluster Survey \citep{ferrarese2012a} obtained with MegaCam \citep{boulade2003a} have an effective resolution of $\sim \SI{50}{\pc}$, prohibiting an analysis of the NSC sizes \citep{ferrarese2020a}.

In a pure dissipationless merger scenario, the steepness of the slope of the cluster may not exceed that of its progenitors \citep{dehnen2005a}.
The slope of the density profiles is determined by evaluating $\mathrm{d} \log_{10} I / \mathrm{d} \log_{10} r$ from \Cref{equ:sersic_profile} and converting to $\mathrm{d} \log_{10} M_{\star} / \mathrm{d} \log_{10} r$ and using the mass-to-light ratio,
\begin{equation}
  \label{equ:nsc_slope}
  \frac{\mathrm{d} \log_{10} \, M_{\star}^{\mathrm{nsc}}}{\mathrm{d} \log_{10} r} = - \frac{\log 10}{r_{\mathrm{eff}^{2}}} \frac{b_{n}}{n} \bigg( \frac{r}{r_{\mathrm{eff}}} \bigg)^{1 / n} \; .
\end{equation}
For the new detections, we find an increase in this slope but the trends are not significant.
Based on a similar trend and a comparison to typical GC densities, \citet{pechetti2020a} concluded that \textit{in-situ} star formation plays a key role in the formation and evolution of NSCs.
For the majority of our clusters, it remains unclear whether \textit{in-situ} star formation contributes to the mass budget at all.
\citet{fahrion2022b} showed that most of the mass fraction of NSCs in similar-mass galaxies comes from old, metal-poor stars but that \textit{in-situ} star formation may still be present.

If low-mass NSC structure argues for GC merging as the primary formation channel, then at the highest NSC masses, we do see some evidence for in-situ formation
The two highest-mass NSCs in our sample are denser than the densest GCs, including Milky Way clusters \citep[e.g.][]{mclaughlin2005a,baumgardt2018b}, and many ultra-compact dwarfs \citep[e.g.][]{norris2014a}.
This hints towards a contribution of \textit{in-situ} star formation, supported by \citet{fahrion2022a} who found that \text{in-situ} star formation gains importance for $\log_{10} \, M_{\star}^{\mathrm{nsc}} / \mathrm{M}_{\odot} \gtrsim 6.5$ and may contribute \SI{50}{\percent} of the NSCs mass.
Whether these objects are a product of repeated GC mergers is plausible but remains unclear as the increase in profile slope may be caused by central \textit{in-situ} star formation.

Combining all arguments, it appears to be clear that there is a fundamental connection between GCs and NSCs in these low-mass galaxies.
Although likely, it remains unclear whether the lowest-mass NSCs are individual GCs, which experienced no merger events, or whether the NSCs are the product of GCs mergers.
At least the two most-massive NSCs in our sample likely experienced \textit{in-situ} star formation, elevating the steepness of their profile slopes and making them denser than any Milky Way GC.

\section{Conclusions}
\label{sec:conclusions}

In this work we presented an analysis of \num{21} newly discovered nuclear star clusters (NSCs) in Local Volume galaxies using \textit{Hubble Space Telescope} imaging data.
We convolved a {\tinytim}-generated point spread function with a S{\'{e}}rsic profile to determine structural parameters.
NSC stellar masses were determined based on integrated photometry in different filters.

The new detections are compact with a typical effective radius $r_{\mathrm{eff}} \lesssim \SI{12}{\pc}$ and populate the lower stellar mass end of the whole NSC population at $M_{\star}^{\mathrm{NSC}} \lesssim \SI{e7}{\Msun}$.
We find that the correlation between $M_{\star}^{\mathrm{nsc}}$ and $r_{\mathrm{eff}}$ breaks down for the low-mass galaxies, as indicated by \citet{georgiev2016a}.
In addition to their compact size, the new detections have typically low- to moderate S{\'{e}}rsic indices ($n \lesssim 6$), which compares to other NSCs in the Local Volume.
The linear relation between the ellipticity and the mass of the clusters break down below $M_{\star}^{\mathrm{gal}} \sim \SI{e6.5}{\Msun}$ where the NSCs have ellipticies of $\epsilon \sim 0.1$.
A comparison to Milky Way globular clusters \citep{harris1996a,baumgardt2018b} reveals that most of the newly detected NSCs have similar ellipticity, effective radius, and stellar mass, corroborating a relation between both types of clusters.

NSCs are the densest stellar systems \citep[e.g.][]{walcher2005a,norris2014a,neumayer2020a} and we find central surface brightness values ranging between $\sim18$ and $\sim \SI{12}{\mag\per\arcsec\squared}$ in the \textit{F814W} band, corresponding to central surface masses of $\sim 3.2$ and \SI{6.2}{\Msun\per\parsec\squared}, respectively.
We find that both the surface brightness and stellar mass profiles correlate with both the NSC and host galaxy stellar mass.
Furthermore, the slope of the profiles evaluated at their effective radii weakly correlates with both the NSC and host galaxy stellar mass.
A similar trend for three dimensional slope values was observed by \citet{pechetti2020a} for more massive NSCs.
Our data reveal that this trend continues down to the lowest-mass nucleated galaxies.

Similar to the surface brightness profiles, the average surface mass density within the effective radius correlates with NSC stellar mass as well.
A linear fit reveals that some denser and more massive NSCs follow the same trend, albeit their distribution widens and flattens towards higher masses.
Comparing to Milky Way globular clusters, we find that about \SI{65}{\percent} fall above the best-fit relation.
Again, most of the lowest-mass NSCs coincide with the distribution of Milky Way globular clusters.

We investigated the scaling relation of NSC versus host galaxy mass.
A linear fit revealed that the nucleated dwarfs in a field environment have a steeper relationship ($\alpha = 0.82^{+0.08}_{-0.08}$) than dwarfs in the core of the Virgo galaxy cluster ($\alpha = 0.55^{+0.06}_{-0.05}$; \citealp{sanchez-janssen2019a}).
However, forcing $M_{\star}^{\mathrm{gal}} \geq \SI{e6.5}{\Msun}$ for the fit results in a relationship with a steepness comparable to the value for dwarfs in the field environment ($\alpha = 0.70^{+0.08}_{-0.07}$).
Therefore, the environmental dependence in the $M_{\star}^{\mathrm{nsc}}$-$M_{\star}^{\mathrm{nsc}}$ relation is caused by the lowest-mass nucleated galaxies.

Our results reinforce the connection between globular clusters and nuclear star clusters.
They also corroborate other studies in that globular cluster migration is the main formation mechanism in dwarf galaxies and that \textit{in-situ} star formation gains importance with increasing mass \citep[e.g.][]{neumayer2020a}.

We find a clear environmental dependence, such that in low-mass galaxies, the NSCs are fractionally more massive in denser environments. We argue this extra mass is most likely explained by a larger pool of available GCs for mergers, or even just for becoming the NSC. On the flip side, the high stellar density of our two most massive NSCs suggest that in-situ formation, rather than merging, dominated their growth.
This interpretation fits well with other recent research, which shows that the \textit{in-situ} fraction of a nuclear star cluster increases with increasing stellar mass \citep{fahrion2022a,fahrion2022b}.
Our data cannot reveal whether there also exists an environmental dependence in the correlation between the NSCs' \textit{in-situ} fraction and stellar mass.

\section*{Acknowledgements}
The authors thank the editor and referee for constructive feedback.
A.C.S.\ acknowledges support from NSF grant AST-2108180.
This research is based on observations with the NASA/ESA \textit{Hubble Space Telescope}, obtained at the Space Telescope Science Institute, which is operated by AURA, Inc., under NASA contract NAS5-26555.
This research has made use of the HyperLEDA data base \citep{makarov2014a}, the SIMBAD data base \citep{wegner2000a}, and NASA's Astrophysics Data System (ADS).
We made use of \textsc{Astropy} \citep{astropy2013a,astropy2018a}, \textsc{NumPy} \citep{harris2020a}, \textsc{Dustmaps} \citep{green2018a}, \textsc{Matplotlib} \citep{hunter2007a}, \textsc{Scipy} \citep{virtanen2020a}, {\imfit} \citep{erwin2015a}, and {\tinytim} \citep{krist1993a,krist1995a}.


\section*{Data availability}

The data underlying this article are available in the article and in its online supplementary material.




\bibliographystyle{mnras}
\bibliography{references.bib} 




\appendix

\section{Assessing uncertainties}
\label{sec:assessing_uncertainties}

In this section we discuss statistical and systematic uncertainties and how we determined them.
If applicable, the final $1\sigma$ uncertainties in the data tables consist of the sum of the quadratic statistical and systematic uncertainties.

\subsection{Statistical uncertainties}
\label{subsubsec:statistical_uncertainties}

For each fit, we determine the statistical uncertainties via bootstrapping.
During each iteration of bootstrapping {\imfit} generates a new data array where pixel values are resampled from the original data.
The new data array is then fit using Levenberg-Marquardt minimisation to speed up the fit.
We chose \num{500} bootstrap iterations which resulted in a good-enough sampling of the confidence intervals; increasing the value to \num{2500} iterations did not change the results.

The quoted uncertainties were determined by the $1\sigma$ distribution of the bootstrap results.
However, to determine photometric parameter values and to convert the effective radius from pixels to parsecs, the uncertainties needed to be propagated forward.
In the case of apparent magnitudes, we used the bootstrap distributions of all required parameter values to determine the total intensity of the NSC (\textit{cf}.~\Cref{equ:nsc_intensity}).
The uncertainty on the zeropoint magnitude is small\footnote{typically $\mathcal{O}(\SI{e-3}{\mag})$.} compared to the uncertainty of the instrumental magnitude and was not taken into account.
For the determination of colours, we used Gaussian error propagation by assuming that the distributions of apparent magnitudes follow a Gaussian distribution.
We used the larger uncertainty of the asymmetric parameter distribution as the symmetric uncertainty of the assumed Gaussian-like distribution.
For the apparent magnitudes, this choice seems to be justified, as shown by the symmetry of the uncertainties of the apparent magnitudes in \Cref{tab:nsc_parameters}.
Afterwards, we determined absolute magnitudes and stellar masses via Gaussian error propagation.
The same scheme was applied to transform effective radii from pixels to parsecs.

\subsection{Systematic uncertainties}
\label{subsubsec:systematic_uncertainties}

To quantify systematic uncertainties in our work, we conducted various tests involving the choice of model functions and the programs {\astrodrizzle} and {\tinytim}.
We will additionally discuss the correlation between the S{\'{e}}rsic index and the effective radius and the induced uncertainty by fixating the index in some of our fits.
All fits were performed with {\imfit} and are independent of the chosen solver or fit statistic.
Unless otherwise stated, we chose the data of NGC{\,}2337 in the ACS/WFC $F814W$ band.

\subsubsection{Model functions}
\label{subsubsec:model_functions}

We assumed that the NSCs can be represented well by a single S{\'{e}}rsic function but this choice is rather arbitrary.
Complex substructures may be present in extragalactic NSCs but typically are unresolved given their distances and subsequent angular sizes on the {\hst} instruments.
Nevertheless, in some cases individual stars (e.g. [KK2000]{\,}03) and extended emission around the NSC can be seen which are not well represented by a single S{\'{e}}rsic profile.

We repeated most fits using different model functions for the NSC.
We chose a single King profile \citep{king1962a,king1966a}, a combination of a S{\'{e}}rsic profile and a point source, two S{\'{e}}rsic profiles where the second profile fits the extended emission, and two S{\'{e}}rsic profiles in combination with a point source.
The addition of a point source to the fits was tested for all NSCs but did not yield different structural parameters.
In most cases the intensity of the point source was insignificant compared to the intensity of the S{\'{e}}rsic profile at the effective radius, thus not adding significant flux to the total apparent magnitude.

Instead of using a S{\'{e}}rsic profile, we used a classical King profile to fit the NSC of LeG{\,}09 in the ACS/WFC $F814W$ band.
The boundaries for the core and tidal radii were set to $[0.01, 10]$ and $[0.01, 50]$.
Fitting LeG{\,}09 with a S{\'{e}}rsic profiles and using \num{500} bootstrap iterations resulted in $r_{\mathrm{eff}}^{\textrm{s{\'{e}}rsic}} = 3.19^{+0.12}_{-0.24}$ pixel.
Repeating the fit with a King profile and using the transformation from \citet{georgiev2019a}, which connects the core and tidal radii of the King profile with the effective radius, results in $r_{\mathrm{eff}}^{\mathrm{king}} = 3.05^{+0.06}_{-0.07}$ pixel.
This value lies within the $1\sigma$ statistical uncertainty of the previous fit.
Additionally, we added a second S{\'{e}}rsic profile to the fit resulting in a similar result:
the flux of the fit with two profiles had a higher flux by \SI{\sim 1.6}{\percent} which corresponds to a difference in magnitude of \SI{\sim 0.007}{\mag} which is far below the statistical uncertainty.
Therefore, based on this test, we conclude that the choice of a single S{\'{e}}rsic profile seems to be justified and that the systematic uncertainties induced by this choice are negligible.

Our fits also assume that a constant offset accounts for the underlying light profile (background and galaxy).
The only two exceptions are UGC{\,}01104 and UGC{\,}09660 where a second S{\'{e}}rsic profile needed to be added for the galaxy component.
Not including this second profile leads to a fit of the underlying galaxy profile and not the NSC.
In all other cases, the assumption of local flatness may not be justified, especially for high surface brightness galaxies with complex central structures.
For low surface brightness galaxies, $n \lesssim 1$ \citep[e.g.][]{carlsten2022a} and $r_{\mathrm{eff}}^{\mathrm{gal}} \gg r_{\mathrm{eff}}^{\mathrm{nsc}}$, and the assumption of local flatness seems justified.
Additionally, we only consider the proximity of the NSC where the side length of the fitting area (e.g. \SI{100}{\pixels}) is considerably smaller than $r_{\mathrm{eff}}^{\mathrm{gal}}$.
Also, as mentioned in \Cref{subsec:image_processing}, changing the extent of the fitting region does not change the fit results.

\begin{figure*}
  \centering
  \includegraphics[width=\textwidth]{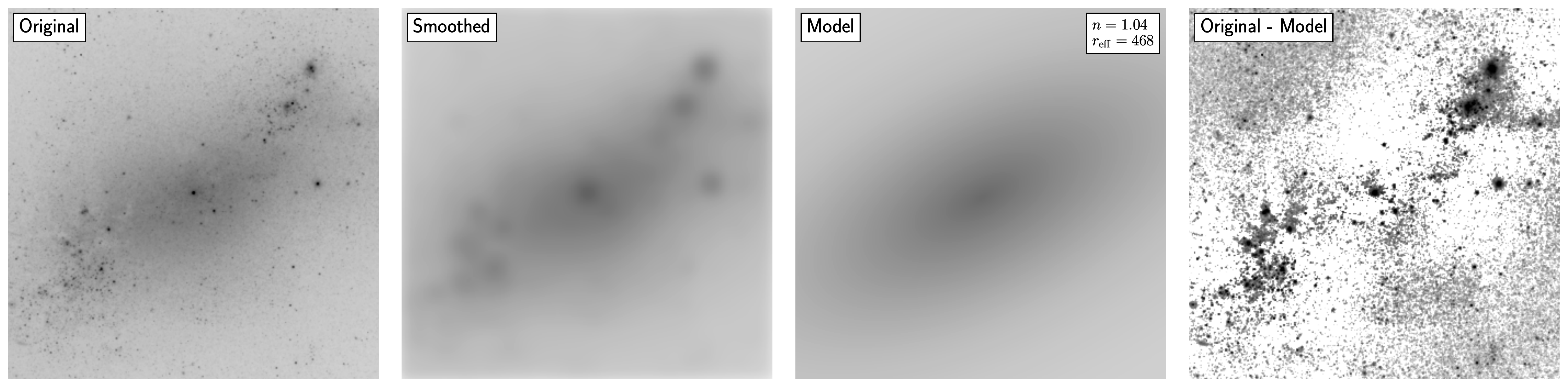}
  \caption{%
    \textit{First panel}: NGC{\,}2337 ACS/WFC $F814W$ data product centred on the nuclear star cluster.
    The square region has side length \SI{1000}{\pixels} which corresponds to \SI{\sim 2700}{\parsec}.
    \textit{Second panel}: Smoothed version of the data shown in the first panel.
    We smooth the data using a two dimensional Gaussian kernel with a standard deviation of $\sigma = \SI{21}{\pixels}$.
    \textit{Third panel}: Fit to the smoothed data product shown in the second panel.
    We approximate the bar component with a S{\'{e}}rsic profile.
    The fit was performed with {\imfit} and the resulting S{\'{e}}rsic index $n$ and effective radius $r_{\mathrm{eff}}$ are indicated in the top right corner.
    \textit{Fourth panel}: Residual map showing the difference between the science data (first panel) and the model of the bar (third panel).
  }
  \label{fig:background_component}
\end{figure*}

To test the systematic uncertainty induced by assuming local flatness, we considered NGC{\,}2337 which features a prominent bar (\textit{cf}.~\Cref{fig:imaging_example}) and, thus, should be the most affected galaxy in the sample\footnote{All other galaxies do not show such a bar and could be approximated by a single S{\'{e}}rsic profile.}.
As shown in \Cref{fig:background_component}, we selected a squared region of \SI{1000}{\pixels} centered on the NSC and applied a 2D Gaussian smoothing kernel with a standard deviation of \SI{21}{\pixels} to determine reliable parameter estimates.
Point or compact sources do not drastically influence the fit due to the applied smoothing.
Approximating the bar component with a S{\'{e}}rsic profile yields $n \sim 1.0$ and $r_{\mathrm{eff}}^{\mathrm{bar}} \sim \SI{470}{\pixel}$ (third panel in \Cref{fig:background_component}).
We repeated the fit of the NSC on the original science product (i.e.\ without applying the smoothing kernel) while keeping all structural parameters for the S{\'{e}}rsic profile describing the bar component fixed.
The fit resulted in $r_{\mathrm{eff}} = 1.01^{+0.12}_{-0.03}$ for the NSC, whereas we found $r_{\mathrm{eff}} = 1.11^{+0.12}_{-0.08}$ pixel in the fit without accounting for the bar.
The difference in magnitude is \SI{\sim 0.02}{\mag} and is smaller than the statistical uncertainty.
Given these values we conclude that our assumption of local flatness is justified.

Finally, the structure of the underlying light distribution of the galaxy could depend on the filter used for fitting.
We evaluated this potential issue by following \citet{pechetti2020a} who fit the NSC in the reddest filter and kept the structural parameters fixed in the bluer filters.
For NGC{\,}2337 we first fit the ACS/WFC $F814W$ data followed by the $F606W$ data.
The fit on the $F606W$ data with the structural parameters of the $F814W$ yielded a difference in magnitude of \SI{\sim 0.12}{\mag} which is larger than the statistical uncertainty on the magnitude.
However, this magnitude is only used for determining the colour of the NSC and eventually the stellar mass where the uncertainty budget is dominated by the uncertainty on the stellar mass-to-light ratio (\SI{0.3}{\dex}).
Furthermore, while the galactic background might change, the structural properties of the NSC, such as the S{\'{e}}rsic index or effective radius, may change as well given the complexity of NSCs and potentially radially varying stellar populations (e.g.\ \citealp{georgiev2014a} and \Cref{subsubsec:wavelength_dependence}).
Finally, as discussed further in \Cref{subsubsec:fixation_of_sersic_indices}, the S{\'{e}}rsic index is unknown for this source and may change as a function of wavelength as well.
In conclusion, we note that for the apparent magnitude in the $F606W$ the found systematic uncertainty appears larger than the statistical uncertainty, but variations in NSC structure could be the origin of these differences.
We decide to not follow the approach by \citet{pechetti2020a} and fit all filters independently of each other.

\subsubsection{Tests on {\astrodrizzle} \& {\tinytim} using simulated data}
\label{subsubsec:astrodrizzle_and_tinytim}

Other systematic uncertainties could be induced by either {\astrodrizzle} or {\tinytim}.
To test the chosen settings for both programs, we generated mock NSC data using the \texttt{makeimage} function of {\imfit}.

\begin{figure*}
  \centering
  \includegraphics[width=0.8\textwidth]{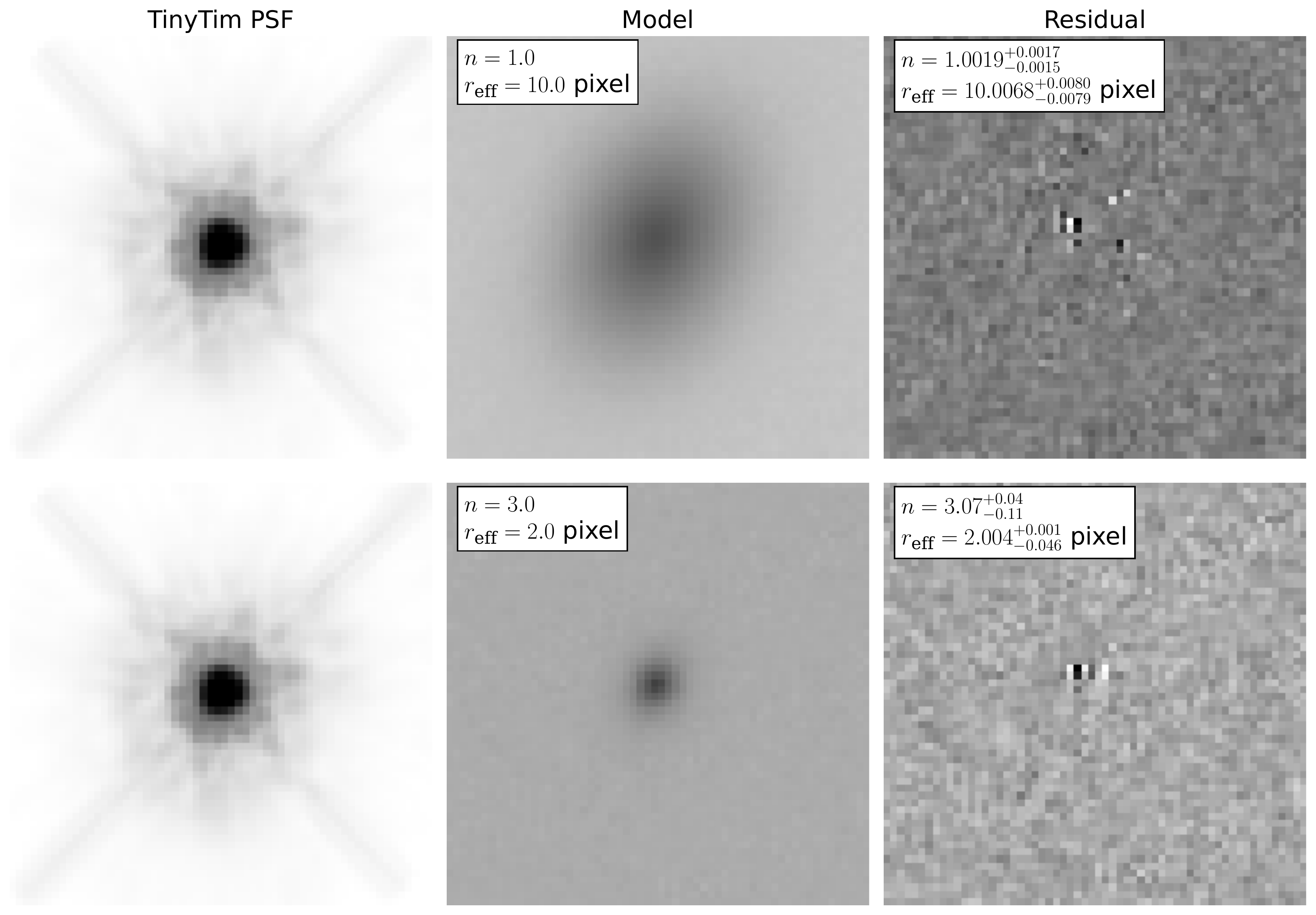}
  \caption{%
    \textit{Left panels}: {\tinytim}-generated PSFs using a \texttt{G2V} star as stellar template.
    The PSFs were superimposed on a normalised version of the science data of NGC{\,}2337, processed with {\astrodrizzle} and extracted from the output image.
    Both PSFs are identical.
    \textit{Middle panels}: 2D S{\'{e}}rsic profiles which have been convolved with the PSFs shown in the left panels.
    In addition, a flat background was added where the pixel values were randomly drawn from a Gaussian distribution.
    The top panel shows an extended profile whereas the bottom one is more compact and has a steeper centre, as indicated by the parameter values.
    The data processing with {\astrodrizzle} is equal to the approach used for the PSFs shown in the left panels.
    \textit{Right panels}: Residual maps from fitting the S{\'{e}}rsic models (middle panels) using the PSFs shown in the left panels with {\imfit}.
    The structural parameters of the fit should equal the values used to generate the S{\'{e}}rsic models and are indicated in the central pictures.
  }
  \label{fig:sys_ad_ttpsf}
\end{figure*}

Simulated NSCs were created by convolving a S{\'{e}}rsic profile with a {\tinytim}-generated PSF.
We added a flat background component whose values were randomly drawn from a Gaussian distribution.
To test the influence of the settings of {\tinytim}, we fed this model and the PSF to {\imfit} and tried to recover the initial S{\'{e}}rsic indices and effective radii.
To test the influence of {\astrodrizzle}, we took the science data of NGC{\,}2337 and normalised it.
We then superimposed the simulated NSC at the location of the NSCs on each exposure and performed {\astrodrizzle}.
Afterwards, the simulated NSC was obtained from the output image of {\astrodrizzle}.
The {\tinytim}-generated PSF was processed in the same way.
The output data of {\astrodrizzle} were fed to {\imfit} where we tried to recover the initial S{\'{e}}rsic index and the effective radius.
We repeated the fits for different S{\'{e}}rsic indices and effective radii starting from $(n = 1, r_{\mathrm{eff}} = \SI{10}{\pixel})$ and going to $(n = 3, r_{\mathrm{eff}} = \SI{2}{\pixel})$ in steps of $\delta n = 1$ and $\delta r_{\mathrm{eff}} = \SI{-1}{\pixel}$ (i.e.\ \num{27} different settings).
Two examples for the PSFs and the simulated NSCs are shown in the two left columns of \Cref{fig:sys_ad_ttpsf}.
The middle column shows two simulated NSCs convolved with the PSF and the right panels show the residual maps, including the recovered structural parameters and their uncertainties, as determined via \num{500} bootstrap iterations.

If both the simulated NSC and the {\tinytim}-generated PSFs were not processed by {\astrodrizzle}, we recovered the initial structural parameter values for all combinations of $n$ and $r_{\mathrm{eff}}$ to high precision.
Once we include {\astrodrizzle} for both the simulated NSC and the PSF, while using the same settings as for the science data, the structural parameters are recovered within the $1\sigma$ interval.
The agreement with the initial parameter values is best for large effective radii and small S{\'{e}}rsic indices and becomes worse with more compact sources and steep profiles.

The recovered parameters became worse once we did not process the PSF with {\astrodrizzle}.
In this case the PSFs were directly taken from {\tinytim} and rotated according to the orientation of the {\astrodrizzle} output.
The uncertainty of the fit became larger and, in the case of a compact source with a steep inner slope, we were unable to recover the S{\'{e}}rsic index within the $1\sigma$ uncertainty distribution.
Therefore, a significant systematic uncertainty is induced if the {\tinytim}-generated PSF is not processed in the same way as the science data.

\begin{figure}
  \centering
  \includegraphics[width=\columnwidth]{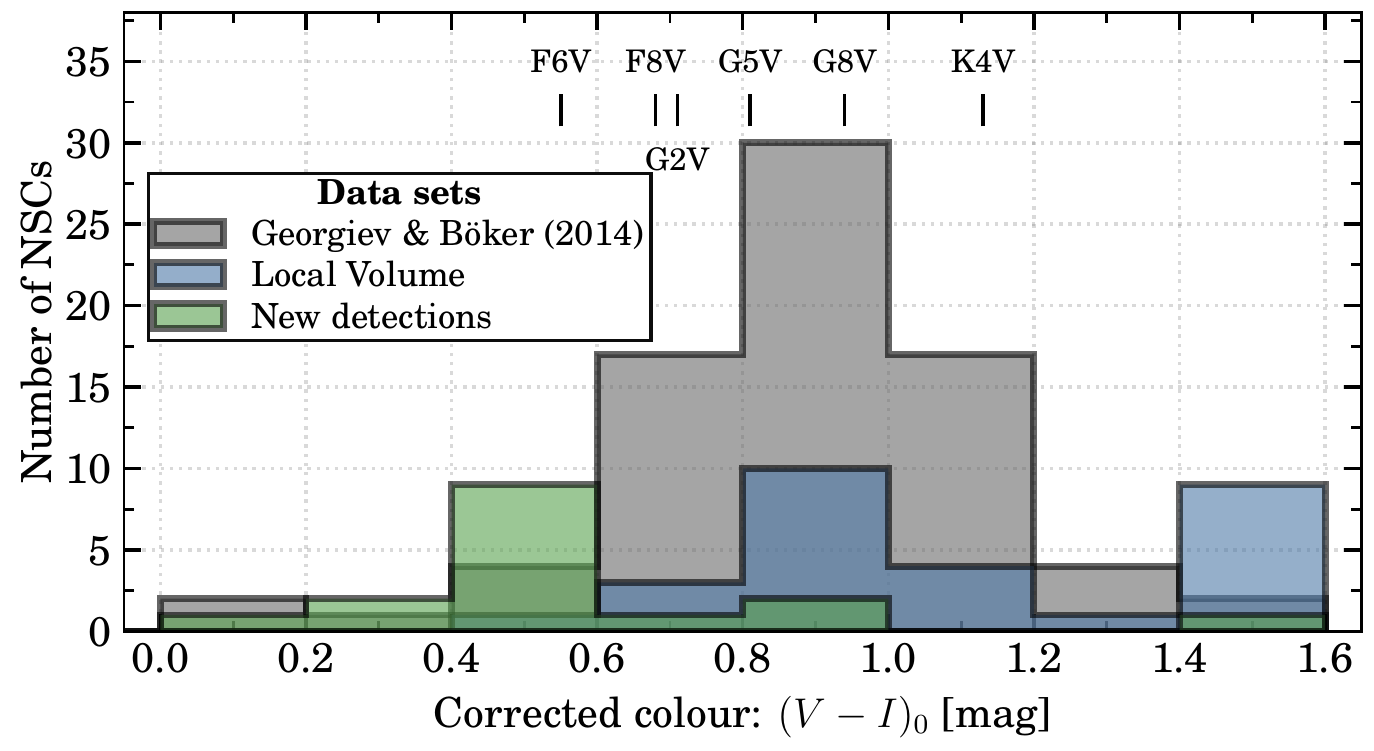}
  \caption{%
    Nuclear star cluster $(V - I)_{0}$ colour.
    We compare our new detections (green) to other Local Volume data (blue) and the data set of \citet{georgiev2014a} for nuclear star clusters in massive late-type field galaxies (gray).
    In addition, we highlight the colour of the different stellar templates tested for the synthetic point spread function.
    For the analysis in the main part of the paper, the template of a \texttt{G2V} star is used.
  }
  \label{fig:nsc_colours}
\end{figure}

We also tested settings related to {\tinytim}.
We generated different PSFs assuming stellar templates ranging from \texttt{F6V} ($V - I = \SI{0.55}{\mag}$) to \texttt{K4V} ($V - I = \SI{1.13}{\mag}$) which covers the colour-range of typical NSCs (\textit{cf}.~\Cref{fig:nsc_colours}).
Using these different PSFs on various science data yielded no significant differences in the resulting parameter values.

In addition, we evaluated whether the accuracy of placing PSFs onto normalised science data is an issue.
More specifically, taking {\tinytim}-generated PSFs and superimposing them onto the normalised science data results in an accuracy of \SI{\pm 1}{\pixel}.
Therefore, we generated subsampled PSF (with subsampling factor ten), superimposed them on normalised single exposures (ACS/WFC data), processed the data with {\astrodrizzle}, resized the PSFs to the resolution of the science data, and applied the charge diffusion kernel.
This approach should yield an accuracy of \SI{\sim 0.1}{\pixel}.
After fitting a few NSCs, we again found no significant differences and conclude that both the settings chosen in {\tinytim} and the uncertainty induced by placing PSFs onto normalised science exposures are insignificant.

Finally, we checked the modified {\astrodrizzle} parameters `pixel fraction' and `resolution'.
As briefly discussed in \Cref{subsec:image_processing}, we chose a value of \num{0.75} for the pixel fraction and increased the final resolution according to the extent of the theoretical PSF for the ACS/WFC products.
We conducted tests where we changed both the pixel fraction (between \num{0.5} and \num{1.0}) and the final resolution [between \SI{0.035}{\arcsec\per\pixel} and \SI{0.05}{\arcsec\per\pixel} (original resolution)] but found no difference in the resulting parameter values.
However, artifacts appeared in the weight maps of the data when choosing a low value for either the pixel fraction and resolution which indicate that {\astrodrizzle} cannot find input pixels from the individual exposures to generate a pixel value on the output grid.
We verified that the different {\astrodrizzle} settings do not change the recovered structural parameter values of our mock data.

In conclusion, we cannot find significant systematic uncertainties induced by our approach.
Note that systematic uncertainties can become significant once the PSF is not processed in the same way as the science data.

\subsubsection{Fixation of S{\'{e}}rsic indices}
\label{subsubsec:fixation_of_sersic_indices}

\begin{figure}
  \centering
  \includegraphics[width=\columnwidth]{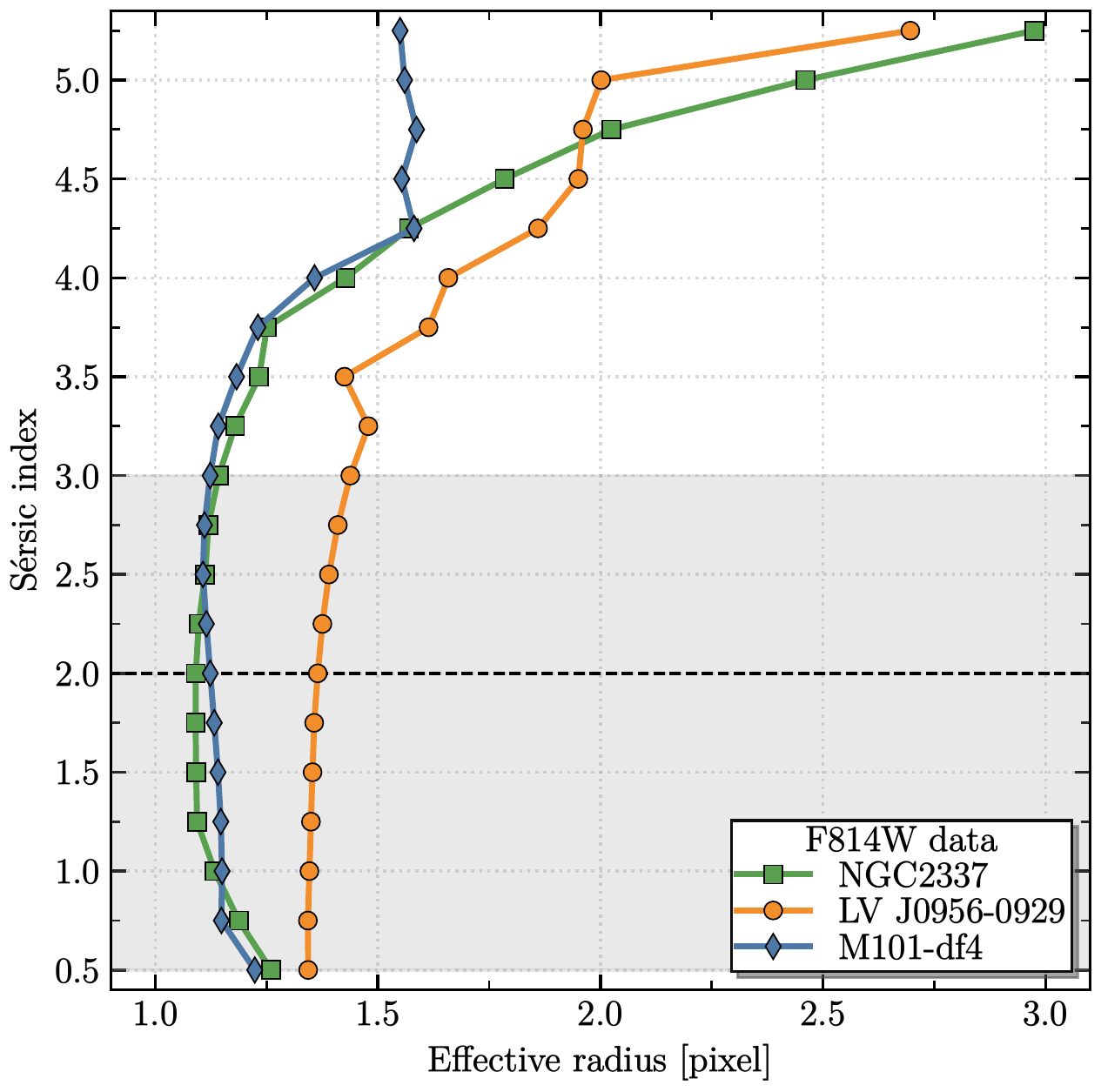}
  \caption{%
    Fit results of three different nuclear star clusters (NGC{\,}2337: green, LV{\,}J0956-0929: orange, M101-df4: blue) with fixed S{\'{e}}rsic indices.
    The dashed horizontal line shows the S{\'{e}}rsic value $n = 2$ used to obtain an approximate value for the effective radii of the NSCs.
    The gray shaded area shows the range of S{\'{e}}rsic values which we consider to be reasonable given the indices of quality zero fits (i.e.\ `good' fits) presented by \citet{pechetti2020a}.
  }
  \label{fig:sys_sersic_r_eff}
\end{figure}

As discussed in \Cref{subsec:fitting_procedure}, the S{\'{e}}rsic index of a few NSCs diverges.
To be able to approximate the effective radius of the NSCs, we fixed the index to a value of $n = 2$.
This value roughly equals the median value of the quality zero fits of the data set of \citet{pechetti2020a}.
We investigated the induced systematic uncertainty of this choice in \Cref{fig:sys_sersic_r_eff} where we show the S{\'{e}}rsic index versus effective radius.
The plot shows three NSCs for which we repeated the fit with varying S{\'{e}}rsic indices.
You can see that the effective radius is only slightly affected by the choice of S{\'{e}}rsic index between $n = 0.5$ and $n = 3.5$.
At higher values of $n$ the effective radius increases and appears to diverge towards higher values.
The only exception is M101-df4 for which $r_{\mathrm{eff}}$ appears to remain constant.

The figure shows that there exists a systematic uncertainty induced by fixing the S{\'{e}}rsic index to a value of $n = 2$.
Therefore, we determine the largest differences between effective radius between $n = 0.5$ and $n = 3$ and add this value in quadrature to the larger statistical uncertainty obtained from bootstrap iterations.
If the `true' S{\'{e}}rsic index is larger than $n = 3$, our quoted effective radii become systematically too small, but as we show in \Cref{subsec:structural_properties}, this issue does not affect our results.

Due to the choice of $n = 2$, the apparent magnitudes of the NSCs are affected as well.
In our tests the difference in magnitude is typically $\delta m \sim \SI{0.1}{\mag}$ when setting $n = 0.5$ and $n = 3$.
Therefore, for the cases where we set $n = 2$ we add in quadrature to the statistical uncertainty the statistical uncertainty \SI{0.1}{\mag}.

Finally, we tested the effect of fixing $n = 2$ for the NSCs where the index did not diverge in the fits.
Repeating the fits and using \num{500} bootstrap iterations we find typical differences of $\delta r_{\mathrm{eff}} \lesssim \SI{5}{\percent}$ which is comparable to the statistical uncertainty

\section{Data tables}
\label{sec:data_tables}

Here we present the data tables underlying this article.
\Cref{tab:available_data} gives an overview of the galaxies hosting the newly discovered NSCs and their available {\hst} data.
Galaxy properties are adapted from \citet{hoyer2021a} and raw images, containing exposure information, are taken from the {\hla}.
\Cref{tab:nsc_parameters} presents properties of the newly discovered NSCs.
\Cref{tab:lit_nsc_parameters} gives the parameters of other NSCs in the Local Volume and in \Cref{tab:lit_georgiev2014} we present the NSC stellar mass estimates of the data sample of \citet{georgiev2014a}.
\begin{table*}
  \scriptsize
  \caption{%
    List of \num{21} galaxies whose nuclear star clusters are new discoveries, sorted by descending galactic stellar mass.
    Galactic parameters (columns 2-5) are taken from \citet{hoyer2021a}.
    The Proposal IDs and exposure times $t_{\mathrm{exp.}}$ are taken from the data products.
    For the ACS, we use an online calculator to obtain the zeropoint magnitudes: \url{https://acszeropoints.stsci.edu/}
    For the WFPC2, we adopt the zeropoint magnitudes from the instrument's manual \citep{wfpc2_manual}.
    For the WFC3, we use an online data base: \url{https://www.stsci.edu/hst/instrumentation/wfc3/data-analysis/photometric-calibration/uvis-photometric-calibration}.
  }
  \begin{center}
    \begin{threeparttable}
      \begin{tabular}{%
        l
        S[table-format=3.5]
        S[table-format=2.5]
        l
        S[table-format=1.3]
        l
        l
        S[table-format=5.0]
        S[table-format=2.3]
        S[table-format=4.0]
        S[table-format=1.4]
      }
        \toprule
        \multicolumn{1}{c}{Name} & {RA} & {DE}  & \multicolumn{1}{c}{dm} & {$\log_{10} \, M_{\star}$} & \multicolumn{1}{c}{Instrument} & \multicolumn{1}{c}{Filter} & {Proposal ID} & {$m_{\mathrm{Inst.}}$} & {$t_{\mathrm{exp.}}$} & {pixel scale}\\
        \cmidrule(lr){2-2}
        \cmidrule(lr){3-3}
        \cmidrule(lr){4-4}
        \cmidrule(lr){5-5}
        \cmidrule(lr){9-9}
        \cmidrule(lr){10-10}
        \cmidrule(lr){11-11}
         & {[deg]} & {[deg]} & \multicolumn{1}{c}{[mag]} & {[M$_{\odot}$]} & {} & {} & {} & {[VEGAmag]} & {[s]} & {[\si{\arcsec\per\pixel}]}\\
        \midrule

        \multirow{2}{*}{NGC\,2337}             & {\multirow{2}{*}{105.55667}} &  {\multirow{2}{*}{44.45694}} & {\multirow{2}{*}{\num{30.37 \pm 0.49}}} & {\multirow{2}{*}{\num{8.97 \pm 0.34}}} & \multirow{2}{*}{ACS WFC}& {$F814W$} & 13442 & 25.508 & 1000 & \num{0.0472} \\
                                       &                              &                              &                                         &                                        &                         & {$F606W$} & 13442 & 26.395 & 1000 & \num{0.0415} \\

\addlinespace
\multirow{2}{*}{LV\,J0956-0929}        & {\multirow{2}{*}{149.15667}} &  {\multirow{2}{*}{-9.48639}} & {\multirow{2}{*}{\num{29.86 \pm 0.11}}} & {\multirow{2}{*}{\num{8.29 \pm 0.43}}} & \multirow{2}{*}{ACS WFC} & {$F814W$} & 12546 & 25.512 & 900 & \num{0.0472} \\
                                       &                              &                              &                                         &                                        &                          & {$F606W$} & 12546 & 26.398 & 900 & \num{0.0415} \\

\addlinespace
\multirow{2}{*}{[KK2000]\,03}          &  {\multirow{2}{*}{36.17792}} & {\multirow{2}{*}{-73.51278}} & {\multirow{2}{*}{\num{26.50 \pm 0.09}}} & {\multirow{2}{*}{\num{8.16 \pm 0.03}}} & \multirow{2}{*}{ACS WFC} & {$F814W$} & 13442 & 25.510 & 1200 & \num{0.0472} \\
                                       &                              &                              &                                         &                                        &                          & {$F606W$} & 13442 & 26.396 & 1200 & \num{0.0415} \\

\addlinespace
\multirow{2}{*}{UGC\,09660}            & {\multirow{2}{*}{235.28875}} &  {\multirow{2}{*}{44.69806}} & {\multirow{2}{*}{\num{30.16 \pm 0.12}}} & {\multirow{2}{*}{\num{8.16 \pm 0.32}}} & \multirow{2}{*}{ACS WFC} & {$F814W$} & 13442 & 25.510 & 1000 & \num{0.0472} \\
                                       &                              &                              &                                         &                                        &                          & {$F606W$} & 13442 & 26.396 & 1000 & \num{0.0415} \\

\addlinespace
\multirow{2}{*}{UGC\,04998}            & {\multirow{2}{*}{141.30042}} &  {\multirow{2}{*}{68.38306}} & {\multirow{2}{*}{\num{28.58 \pm 0.21}}} & {\multirow{2}{*}{\num{8.13 \pm 0.23}}} & \multirow{2}{*}{WFPC2 PC} & {$F814W$} &  8137 & 21.639 & 1500 & \num{0.05} \\
                                       &                              &                              &                                         &                                        &                           & {$F555W$} &  8137 & 22.545 & 1600 & \num{0.05} \\

\addlinespace
\multirow{2}{*}{UGC\,01104}            &  {\multirow{2}{*}{23.17625}} &  {\multirow{2}{*}{18.31583}} & {\multirow{2}{*}{\num{29.39}}}          & {\multirow{2}{*}{\num{8.00 \pm 0.31}}} & \multirow{2}{*}{WFPC2 WF} & {$F814W$} &  9124 & 21.659 & 80  & \num{0.1} \\
                                       &                              &                              &                                         &                                        &                           & {$F300W$} &  9124 & 19.433 & 600 & \num{0.1} \\

\addlinespace
\multirow{2}{*}{LV\,J1205+2813}        & {\multirow{2}{*}{181.39250}} &  {\multirow{2}{*}{28.23222}} & {\multirow{2}{*}{\num{31.45}}}          & {\multirow{2}{*}{\num{7.92 \pm 0.43}}} & \multirow{2}{*}{ACS WFC} & {$F814W$} & 13750 & 25.510 & 1218 & \num{0.0472} \\
                                       &                              &                              &                                         &                                        &                          & {$F606W$} & 13750 & 26.396 & 1000 & \num{0.0415} \\

\addlinespace 
\multirow{2}{*}{DDO\,133}              & {\multirow{2}{*}{188.22083}} &  {\multirow{2}{*}{31.53917}} & {\multirow{2}{*}{\num{28.44 \pm 0.05}}} & {\multirow{2}{*}{\num{7.80 \pm 0.48}}} & \multirow{2}{*}{WFPC2 WF} & {$F814W$} & 10905 & 21.659 & 2200 & \num{0.1} \\
                                       &                              &                              &                                         &                                        &                           & {$F606W$} & 10905 & 22.896 & 2200 & \num{0.1} \\

\addlinespace 
\multirow{2}{*}{UGC\,07242}            & {\multirow{2}{*}{183.53083}} &  {\multirow{2}{*}{66.09222}} & {\multirow{2}{*}{\num{28.68 \pm 0.03}}} & {\multirow{2}{*}{\num{7.75 \pm 0.43}}} & \multirow{2}{*}{ACS WFC}   & {$F814W$} &  9771 & 25.525 & 900  & \num{0.0472} \\
                                       &                              &                              &                                         &                                        &                            & {$F606W$} &  9771 & 26.414 & 1200 & \num{0.0415} \\

\addlinespace
\multirow{2}{*}{PGC\,154449}           & {\multirow{2}{*}{149.28708}} &  {\multirow{2}{*}{-9.26333}} & {\multirow{2}{*}{\num{29.93}}}          & {\multirow{2}{*}{\num{7.70 \pm 0.39}}} & \multirow{2}{*}{ACS WFC} & {$F814W$} & 15922 & 25.507 & 760 & \num{0.0472} \\
                                       &                              &                              &                                         &                                        &                          & {$F606W$} & 15922 & 26.393 & 760 & \num{0.0415} \\

\addlinespace
\multirow{2}{*}{ESO\,553-046}          &  {\multirow{2}{*}{81.77375}} & {\multirow{2}{*}{-20.67806}} & {\multirow{2}{*}{\num{29.13 \pm 0.02}}} & {\multirow{2}{*}{\num{7.68 \pm 0.63}}} & \multirow{2}{*}{ACS WFC} & {$F814W$} & 12546 & 25.512 & 900 & \num{0.0472} \\
                                       &                              &                              &                                         &                                        &                          & {$F606W$} & 12546 & 26.395 & 900 & \num{0.0415} \\

\addlinespace
\multirow{2}{*}{DDO\,084}              & {\multirow{2}{*}{160.67458}} &  {\multirow{2}{*}{34.44889}}& {\multirow{2}{*}{\num{29.99}}}           & {\multirow{2}{*}{\num{7.65 \pm 0.35}}} & \multirow{2}{*}{ACS WFC} & {$F814W$} & 15922 & 25.507 & 760 & \num{0.0472} \\
                                       &                              &                             &                                          &                                        &                          & {$F606W$} & 15922 & 26.393 & 760 & \num{0.0415} \\

\addlinespace
\multirow{2}{*}{BTS\,76}               & {\multirow{2}{*}{179.68375}} &  {\multirow{2}{*}{27.58500}} & {\multirow{2}{*}{\num{30.50}}}          & {\multirow{2}{*}{\num{7.53 \pm 0.16}}} & \multirow{2}{*}{ACS WFC} & {$F814W$} & 14636 & 25.509 & 1030 & \num{0.0472} \\
                                       &                              &                              &                                         &                                        &                          & {$F606W$} & 14636 & 26.395 & 1030 & \num{0.0415} \\

\addlinespace
\multirow{2}{*}{M\,101-df4}            & {\multirow{2}{*}{211.88917}} &  {\multirow{2}{*}{54.71000}} & {\multirow{2}{*}{\num{32.16}}}          & {\multirow{2}{*}{\num{7.29 \pm 0.41}}} & \multirow{2}{*}{ACS WFC} & {$F814W$} & 13682 & 25.510 & 1150 & \num{0.0472} \\
                                       &                              &                              &                                         &                                        &                          & {$F606W$} & 13682 & 26.393 & 1150 & \num{0.0415} \\

\addlinespace
\multirow{2}{*}{NGC\,5011C}            & {\multirow{2}{*}{198.29958}} & {\multirow{2}{*}{-43.26556}} & {\multirow{2}{*}{\num{27.86 \pm 0.02}}} & {\multirow{2}{*}{\num{7.24 \pm 0.53}}} & \multirow{2}{*}{ACS WFC} & {$F814W$} & 12546 & 25.512 & 900 & \num{0.0472} \\
                                       &                              &                              &                                         &                                        &                          & {$F606W$} & 12546 & 26.398 & 900 & \num{0.0415} \\

\addlinespace 
\multirow{2}{*}{LeG\,09}               & {\multirow{2}{*}{160.64417}} &  {\multirow{2}{*}{12.15056}} & {\multirow{2}{*}{\num{30.04 \pm 0.07}}} & {\multirow{2}{*}{\num{7.08 \pm 0.43}}} & \multirow{2}{*}{ACS WFC} & {$F814W$} & 14644 & 25.509 & 1096 & \num{0.0472} \\
                                       &                              &                              &                                         &                                        &                          & {$F606W$} & 14644 & 26.395 & 1026 & \num{0.0415} \\

\addlinespace
\multirow{2}{*}{[KK2000]\,53}          & {\multirow{2}{*}{197.80917}} & {\multirow{2}{*}{-38.90611}} & {\multirow{2}{*}{\num{27.33 \pm 0.07}}} & {\multirow{2}{*}{\num{6.85 \pm 0.44}}} & \multirow{2}{*}{ACS WFC} & {$F814W$} & 13442 & 25.510 & 1000 & \num{0.0472} \\
                                       &                              &                              &                                         &                                        &                          & {$F606W$} & 13442 & 26.397 & 1000 & \num{0.0415} \\

\addlinespace
\multirow{2}{*}{KK\,96}                & {\multirow{2}{*}{162.61292}} &  {\multirow{2}{*}{12.36083}} & {\multirow{2}{*}{\num{30.00 \pm 0.04}}} & {\multirow{2}{*}{\num{6.68 \pm 0.43}}} & \multirow{2}{*}{ACS WFC} & {$F814W$} & 14644 & 25.509 & 1096 & \num{0.0472} \\
                                       &                              &                              &                                         &                                        &                          & {$F606W$} & 14644 & 26.395 & 1026 & \num{0.0415} \\

\addlinespace
\multirow{2}{*}{LV\,J1217+4703}        & {\multirow{2}{*}{184.29208}} &  {\multirow{2}{*}{47.06361}} & {\multirow{2}{*}{\num{30.70 \pm 0.22}}} & {\multirow{2}{*}{\num{6.58 \pm 0.50}}} & \multirow{2}{*}{ACS WFC} & {$F814W$} & 14644 & 25.509 & 1164 & \num{0.0472} \\
                                       &                              &                              &                                         &                                        &                          & {$F606W$} & 14644 & 26.395 & 1094 & \num{0.0415} \\

\addlinespace
\multirow{2}{*}{PGC\,4310323}          & {\multirow{2}{*}{181.37917}} &  {\multirow{2}{*}{31.07611}} & {\multirow{2}{*}{\num{29.04}}}          & {\multirow{2}{*}{\num{6.55 \pm 0.43}}} & \multirow{2}{*}{ACS WFC} & {$F814W$} & 15922 & 25.507 & 760 & \num{0.0472} \\
                                       &                              &                              &                                         &                                        &                          & {$F606W$} & 15922 & 26.393 & 760 & \num{0.0415} \\

\addlinespace
\multirow{1}{*}{dw\,1335-29}            & {\multirow{1}{*}{203.94167}} & {\multirow{1}{*}{-29.70667}} & {\multirow{1}{*}{\num{28.50 \pm 0.21}}} & {\multirow{1}{*}{\num{6.46 \pm 0.43}}} & \multirow{1}{*}{ACS SBC} & {$F150LP$} & 10608 & 20.747 & 5416 & \num{0.03} \\

\bottomrule

      \end{tabular}
    \end{threeparttable}
  \end{center}
  \label{tab:available_data}
\end{table*}
\begin{table*}
  \scriptsize
  \caption{%
    Structural and photometric parameters of the \num{21} newly discovered nuclear star clusters in the Local Volume.
    All values are the median of the parameter distribution after \num{500} bootstrapping iterations.
    The uncertainties give the $1\sigma$ interval.
  }
  \begin{center}
    \begin{threeparttable}
      \begin{tabular}{%
          l
          l
          l
          l
          l
          l
          l
          l
          l
        }
        \toprule
        \multicolumn{1}{c}{Name} & \multicolumn{1}{c}{Filter} & \multicolumn{1}{c}{PA} & \multicolumn{1}{c}{$\epsilon$} & \multicolumn{1}{c}{$n$} & \multicolumn{1}{c}{$r_{\mathrm{eff}}$} & \multicolumn{1}{c}{$m$} & \multicolumn{1}{c}{$M_{\star} / L_{I}$} & \multicolumn{1}{c}{$\log_{10} \, M_{\star}$} \\
        \cmidrule(lr){3-3}
        \cmidrule(lr){6-6}
        \cmidrule(lr){7-7}
        \cmidrule(lr){8-8}
        \cmidrule(lr){9-9}
         & {} & \multicolumn{1}{c}{[deg]} & {} & {} & \multicolumn{1}{c}{[pc]} & \multicolumn{1}{c}{[mag]} & \multicolumn{1}{c}{[M$_{\odot}$ / L$_{\odot}$]} & \multicolumn{1}{c}{[M$_{\odot}$]} \\
        \midrule

        \multirow{2}{*}{NGC\,2337} & {$F814W$\tnote{(b)}} & $97.2_{-6.5}^{+6.3}$ & $0.22_{-0.04}^{+0.05}$ & $2.0$ & $3.00_{-0.99}^{+0.99}$ & $17.56_{-0.06}^{+0.05}$ & \multirow{2}{*}{$0.4 \pm 2.0$} & \multirow{2}{*}{$6.83 \pm 0.86$} \\
                           & {$F606W$\tnote{(b)}} & $90.6_{-5.2}^{+6.3}$ & $0.22_{-0.04}^{+0.05}$ & $2.0$ & $2.91_{-1.01}^{+1.01}$ & $18.40_{-0.01}^{+0.03}$ \\
\addlinespace
\multirow{2}{*}{LV\,J0956-0929} & {$F814W$\tnote{(b)}} & $53_{-20}^{+20}$ & $0.058_{-0.026}^{+0.033}$ & $2.0$ & $2.97_{-0.41}^{+0.41}$ & $18.76_{-0.01}^{+0.01}$ & \multirow{2}{*}{$0.69 \pm 2.0$} & \multirow{2}{*}{$6.0 \pm 1.3$} \\
                                & {$F606W$\tnote{(b)}} & $80_{-39}^{+39}$ & $0.037_{-0.022}^{+0.023}$ & $2.0$ & $2.80_{-0.33}^{+0.33}$ & $19.46_{-0.01}^{+0.02}$ \\
\addlinespace
\multirow{2}{*}{[KK2000]\,03} & {$F814W$} & $181.4_{-3.1}^{+2.6}$ & $0.106_{-0.008}^{+0.009}$ & $1.265_{-0.033}^{+0.036}$ & $3.75_{-0.16}^{+0.16}$ & $17.75_{-0.01}^{+0.01}$ & \multirow{2}{*}{$0.92 \pm 2.0$} & \multirow{2}{*}{$5.13 \pm 0.94$} \\
                              & {$F606W$} & $179.8_{-2.7}^{+2.6}$ & $0.102_{-0.007}^{+0.007}$ & $1.213_{-0.030}^{+0.029}$ & $3.64_{-0.15}^{+0.15}$ & $18.51_{-0.01}^{+0.01}$ \\
\addlinespace
\multirow{2}{*}{UGC\,09660} & {$F814W$\tnote{(b)}} & $123_{-12}^{+13}$ & $0.121_{-0.038}^{+0.045}$ & $2.0$ & $2.32_{-0.57}^{+0.57}$ & $17.89_{-0.02}^{+0.02}$ & \multirow{2}{*}{$1.0 \pm 2.0$} & \multirow{2}{*}{$6.57 \pm 0.84$} \\
                            & {$F606W$\tnote{(b)}} & $127_{-12}^{+11}$ & $0.126_{-0.031}^{+0.039}$ & $2.0$ & $2.9_{-1.3}^{+1.3}$    & $18.65_{-0.01}^{+0.03}$ \\
\addlinespace
\multirow{2}{*}{UGC\,04998} & {$F814W$\tnote{(b),(c)}} & $93_{-30}^{+25}$  & $0.082_{-0.037}^{+0.046}$ & $2.0$ & $1.12_{-0.35}^{+0.35}$ & $18.40_{-0.01}^{+0.03}$ & \multirow{2}{*}{$19 \pm 2$} & \multirow{2}{*}{$6.97 \pm 0.10$} \\
                            & {$F555W$\tnote{(b),(c)}} & $123_{-10}^{+22}$ & $0.080_{-0.040}^{+0.053}$ & $2.0$ & $1.08_{-0.34}^{+0.34}$ & $20.31_{-0.03}^{+0.06}$ \\
\vdots & \\
\bottomrule

      \end{tabular}
      \begin{tablenotes}
        \item[(a)] The structural parameters in this filter were fixed to the values determined via bootstrapping in the other available filter.
        \item[(b)] The S{\'{e}}rsic index is fixed to $n = 2$ to determine an effective radius. All other parameters are unrestricted (\textit{cf}.~\Cref{tab:parameter_boundaries}).
        \item[(c)] The NSC mass estimate is unreliable as the mass-to-light ratio is too large ($M_{\star} / L_{I} \gtrsim 4 \mathrm{M}_{\odot} / \mathrm{L}_{\odot}$).
        \item[(d)] No fit was possible.
      \end{tablenotes}
    \end{threeparttable}
  \end{center}
  \label{tab:nsc_parameters}
\end{table*}
\begin{table*}
  \scriptsize
  \caption{%
    List of compiled properties of known nuclear star clusters (NSCs) in the Local Volume.
    Only NSCs with available structural parameters are included.
  }
  \begin{center}
    \begin{threeparttable}
      \begin{tabular}{%
          l
          l
          l
          l
          l
          l
          l
          l
          l
          l
          l
          l
          l
        }
        \toprule
        \multicolumn{1}{c}{Name} & \multicolumn{1}{c}{PA$_{F606W}$} & \multicolumn{1}{c}{$\epsilon_{F606W}$} & \multicolumn{1}{c}{$n_{F606W}$} & \multicolumn{1}{c}{$r_{\mathrm{eff}, \, F606W}$} & \multicolumn{1}{c}{PA$_{F814W}$} & {\ldots} & \multicolumn{1}{c}{$(V - I)_{0}$} & \multicolumn{1}{c}{$\log_{10} \, M_{\star}$} & \multicolumn{1}{c}{Reference\tnote{(a)}} \\
        \cmidrule(lr){2-2}
        \cmidrule(lr){5-5}
        \cmidrule(lr){6-6}
        \cmidrule(lr){8-8}
        \cmidrule(lr){9-9}
         & \multicolumn{1}{c}{[deg]} & {} & {} & \multicolumn{1}{c}{[pc]} & \multicolumn{1}{c}{[deg]} & {} & \multicolumn{1}{c}{[mag]} & \multicolumn{1}{c}{[M$_{\odot}$]} & {} \\
        \midrule

        {Circinus}       & {-{-}}                & {-{-}}                 & {-{-}} & {-{-}}              & $160.8^{+5.7}_{-5.7}$   & {\ldots} & $1.5$                  & $7.57^{+0.11}_{-0.11}$ & $13$ \\[0.25em]
{DDO{\,}042}     & {-{-}}                & {-{-}}                 & {-{-}} & {-{-}}              & $115.6^{+4.0}_{-9.0}$   & {\ldots} & {-{-}}                 & {-{-}}                 & $7$  \\[0.25em]
{DDO{\,}082}     & $154.8^{+3.7}_{-4.0}$ & $0.15^{+0.01}_{-0.01}$ & {-{-}} & $0.3^{+0.0}_{-0.0}$ & $148.1^{+14.8}_{-13.4}$ & {\ldots} & $0.94^{+0.01}_{-0.01}$ & $5.97^{+0.24}_{-0.24}$ & $7$  \\[0.25em]
{DDO{\,}088}     & {-{-}}                & {-{-}}                 & {-{-}} & {-{-}}              & $34.4^{+0.6}_{-0.1}$    & {\ldots} & $1.09^{+0.01}_{-0.01}$ & {-{-}}                 & $7$  \\[0.25em]
{ESO{\,}059-001} & {-{-}}                & {-{-}}                 & {-{-}} & {-{-}}              & {-{-}}                  & {\ldots} & {-{-}}                 & $6.16$                 & $6$  \\[0.25em]
\multicolumn{1}{c}{\vdots} & \\

\bottomrule

      \end{tabular}
      \begin{tablenotes}
        \item[(a)] References: (1) \citet{baldassare2014a}; (2) \citet{bellazzini2020a}; (3) \citet{calzetti2015a}; (4) \citet{carson2015a}; (5) \citet{crnojevic2016a}; (6) \citet{georgiev2009a}; (7) \citet{georgiev2014a}; (8) \citet{graham2009b}; (9) \citet{kormendy1999a}; (10) \citet{kormendy2010b}; (11) \citet{nguyen2017a}; (12) \citet{nguyen2018a}; (13) \citet{pechetti2020a}; (14) \citet{schoedel2014a}; (15) \citet{seth2006a}
      \end{tablenotes}
    \end{threeparttable}
  \end{center}
  \label{tab:lit_nsc_parameters}
\end{table*}
\begin{table*}
  \scriptsize
  \caption{%
    List of $V$- and $I$-band apparent magnitudes and stellar mass estimates of nuclear star clusters (NSCs) in the galaxy sample of \citet{georgiev2014a}.
    NSC structural parameters are presented in the data tables of \citet{georgiev2014a}.
  }
  \begin{center}
    \begin{threeparttable}
      \begin{tabular}{%
          l
          l
          l
          l
          l
        }
        \toprule
        \multicolumn{1}{c}{Name} & \multicolumn{1}{c}{$m - M$} & \multicolumn{1}{c}{$m_{V}$} & \multicolumn{1}{c}{$m_{I}$} & \multicolumn{1}{c}{$\log_{10} \, M_{\star}$} \\
        \cmidrule(lr){2-2}
        \cmidrule(lr){3-3}
        \cmidrule(lr){4-4}
        \cmidrule(lr){5-5}
         & \multicolumn{1}{c}{[mag]} & \multicolumn{1}{c}{[mag]} & \multicolumn{1}{c}{[mag]} &\multicolumn{1}{c}{[M$_{\odot}$]} \\
        \midrule

        {DDO{\,}078}     & $27.71 \pm 0.03$ & $19.44 \pm 0.01$ & $18.51 \pm 0.01$ & $5.87 \pm 0.25$ \\
{ESO{\,}138-010} & $30.9 \pm 1.1$   & {-{-}}           & {-{-}}           & {-{-}} \\
{ESO{\,}187-051} & $31.32 \pm 0.89$ & {-{-}}           & {-{-}}           & {-{-}} \\
{ESO{\,}202-041} & $30.58 \pm 0.58$ & {-{-}}           & {-{-}}           & {-{-}} \\
{ESO{\,}241-006} & $31.32 \pm 0.87$ & {-{-}}           & {-{-}}           & {-{-}} \\
\vdots \\

\bottomrule

      \end{tabular}
    \end{threeparttable}
  \end{center}
  \label{tab:lit_georgiev2014}
\end{table*}


\bsp	
\label{lastpage}
\end{document}